\documentclass[journal]{IEEEtran}
\normalsize

\ifCLASSINFOpdf

\else

\fi

\usepackage{cite}
\usepackage{graphicx}
\usepackage{psfrag}
\usepackage{subfigure}
\usepackage{url}
\usepackage{stfloats}
\usepackage{amsmath}
\usepackage{array}
\usepackage{amssymb}
\usepackage{makecell}
\usepackage{stfloats}
\usepackage{booktabs}
\usepackage{threeparttable}
\usepackage{multirow}
\usepackage{float}
\usepackage{subeqnarray}
\usepackage{color}
\usepackage{tabu}

\begin{document}

\title{Bounds on the Capacity Region of the\\Optical Intensity Multiple Access Channel}

\author{Jing~Zhou,~\IEEEmembership{Member,~IEEE}, and~Wenyi~Zhang,~\IEEEmembership{Senior~Member,~IEEE}
\thanks
{This work was supported in part by the National Natural Science Foundation of China through Grant 61722114, Key Research Program of Frontier Sciences of CAS (Grant No. QYZDY-SSW-JSC003), Fundamental Research Funds for the Central Universities under Grant WK3500000005. (\emph{Corresponding author: Wenyi Zhang.})

The authors are with CAS Key Laboratory of Wireless-Optical Communications, University of Science and Technology of China, Hefei, China (e-mail: jzee@ustc.edu.cn; wenyizha@ustc.edu.cn).}
}

\maketitle

\begin{abstract}
This paper provides new inner and outer bounds on the capacity region of the optical intensity multiple access channel (OIMAC) with a per-user average- or peak-power constraint.
For the average-power constrained OIMAC, our bounds at high power are asymptotically tight,
thereby characterizing the asymptotic capacity region.
The bounds are extended to the $K$-user OIMAC with an average-power constraint without loss of asymptotic optimality.
For the peak-power constrained OIMAC, at high power, we bound the asymptotic capacity region to within 0.09 bits, and determine the asymptotic capacity region in the symmetric case.
At moderate power, for both types of constraints, the capacity regions are bounded to within fairly small gaps.
\end{abstract}

\IEEEpeerreviewmaketitle

\section{Introduction}
Intensity modulation and direct detection (IM/DD) based optical wireless communications (OWC), such as visible light communications (VLC), has received increasing attention in recent years \cite{EMH,BSTVH}.
In IM/DD systems, information is transmitted by varying the intensity of emitted light, i.e., the optical power transmitted per unit area.
A widely accepted channel model for IM/DD based indoor OWC is the Gaussian optical intensity channel \cite{KB97,VLC},
which captures key properties including nonnegativity of optical intensity, input-independent additive Gaussian noise,\footnote{This is an accurate model when the noise is dominated by high intensity shot noise from ambient light and/or thermal noise at the receiver \cite{KB97}.}
and practical constraints such as limited average and/or peak optical power.
The Gaussian optical intensity channel has been used in studies on coding and modulation design \cite{HK03,H07,TAKBIM12} as well as channel capacity \cite{HK04,LMW09,FH10,CMA16,ZZ17}.

In many indoor OWC applications there are multiple users (or devices) transmitting data simultaneously \cite{EMH,BSTVH}.
To explore fundamental limits of multiuser indoor OWC, capacities of several multiuser optical intensity channels including parallel channels \cite{CRA2}, multiple access channels (MACs) \cite{CAAA}, broadcast channels \cite{CRA1,SR18}, etc., have been studied.
These channels are building blocks of more complex systems of multiuser indoor OWC.

We consider a discrete-time optical intensity multiple access channel (OIMAC) with Gaussian noise.
In \cite{CAAA}, several bounds on the capacity region of the OIMAC have been established where the input of each user is constrained in both its average and its peak power.
Specifically, the inner bounds were obtained by using truncated Gaussian inputs and uniformly-spaced discrete inputs for each user, respectively;
the outer bounds were obtained by known results for single-user optical intensity channels.
By optimizing both types of input distributions numerically with respect to signal-to-noise ratio (SNR), in \cite{CAAA}, the low-SNR capacity region of the OIMAC can be determined accurately.
However, at moderate to high SNR, the gaps left in \cite{CAAA} are still evident, and the high-SNR capacity region of the OIMAC is still unknown.

In this paper, we provide new inner and outer bounds on the capacity region of the OIMAC with a per-user average-power constraint or a per-user peak-power constraint.
For the average-power constrained OIMAC, we derive asymptotically tight inner and outer bounds at high SNR, thereby determining the high-SNR capacity region.
At moderate SNR the bounds are also fairy tight.
Moreover, we extend the bounds to the $K$-user case without loss of asymptotic optimality, and provide some discussions related to system design.
For the peak-power constrained case, at high peak-to-noise ratio (PNR), the asymptotic capacity region of the OIMAC is bounded to within 0.09 bits, and this gap vanishes in the symmetric case;
at moderate PNR, by combining our outer bound and the inner bound based on discrete inputs in \cite{CAAA}, the capacity region is bounded to within a small gap.
Specifically, a key step to our achievability results is utilizing capacity results of two additive noise channels where the noises obey certain maxentropic distributions, namely, the exponential distribution for the average-power constrained case and the uniform distribution for the peak-power constrained case.
In Table I, we provide a summary of the contributions of this paper as well as a comparison with \cite{CAAA}.
\begin{table*}[tbp]
\renewcommand{\cellset}{\renewcommand{\arraystretch}{1.5}}
\centering
\caption{A Comparison Between Our Work and \cite{CAAA}}
\scalebox{1.05}{
\begin{tabular}{c|c|c|c|c|c}
\hline

\multicolumn{3}{c|}{}&\multicolumn{2}{c|}{\makecell[cc]{\textbf{Our Work}}}&\makecell[cc]{\cite{CAAA}} \\\hline

\multicolumn{3}{c|}{\makecell[cc]{\textbf{Per-user Power} \\\textbf{Constraint}}}&\makecell[cc]{Average power:\\$\mathbf E[X_i]\leq\mathcal E_i$} &\makecell[cc]{Peak power:\\$X_i\leq\mathcal A_i$}&\makecell[cc]{Average and peak power:\\$\mathbf E[X_i]\leq\mathcal E_i=\alpha\mathcal A_i$,\\$X_i\leq\mathcal A_i\mspace{30mu}$}\\ \hline

\makecell[cc]{\multirow{2}*{\rotatebox{90}{{\textbf{Bounding Technique}}$\mspace{-20mu}$}}}&\multicolumn{2}{c|}{\makecell[cc]{\textbf{Inner Bound}\\(Input Distributions)}}&\makecell[cc]{Exponential\\$+$ mixed} &\makecell[cc]{Uniform $+$ \\non-uniformly-\\spaced discrete}&\makecell[cc]{Both truncated Gaussian\\or both uniformly-\\spaced discrete}\\\cline{2-6}

~&\multicolumn{2}{c|}{\makecell[cc]{\textbf{Outer Bound}\\(Single-User Channel\\Results Utilized)}}&\makecell[cc]{Sphere-packing \\bound \cite{HK04}} &\makecell[cc]{Duality-based \\bound \cite{McKellips,TKB}}&\makecell[cc]{Duality-based bound\\\cite{LMW09} and sphere-packing\\bound \cite{CMA16}}\\\hline

\makecell[cc]{\multirow{3}*{\rotatebox{90}{\textbf{Performance of Bounds}$\mspace{15mu}$}}}&\makecell[cc]{\multirow{2}*{\rotatebox{90}{\textbf{Two-User Channel}$\mspace{-26mu}$}}}&\makecell[cc]{\textbf{High SNR} \\\textbf{Asymptotics}}&\makecell[cc]{Tight} &\makecell[cc]{Within 0.09 bits;$^{*}$\\Tight for symme-\\tric constraint}&\makecell[cc]{Within 1.53 bits;$^{*}$\\Within 1.01 bits per user$^{**}$}\\\cline{3-6}

{}&{}&\makecell[cc]{\textbf{Low SNR} \\\textbf{Asymptotics}}&\multicolumn{2}{c|}{Not considered} &\makecell[cc]{Tight}\\\cline{2-6}

{}&\multicolumn{2}{c|}{\makecell[cc]{$K$-\textbf{User Channel}\\\textbf{Asymptotics}}}&\makecell[cc]{High SNR:\\Tight} &\makecell[cc]{Not considered}&\makecell[cc]{High SNR: Constant-gap\\result for each $K$;$^{*}$\\Within 1.01 bits per user.$^{**}$\\Low SNR: Tight}\\\hline

\end{tabular}}

\parbox{1.4\columnwidth}{\small%
\vspace{1eX}
\footnotesize
$^*$For fair comparison, the gap here is with respect to \emph{sum rate}. For a $K$-user ($K\ge2$) OIMAC, the asymptotic gap between outer and inner bounds in \cite{CAAA} is upper bounded by $\frac{1}{2}\log\frac{9(1+\epsilon_1)^2eK}{2\pi}+K\epsilon_2$ nats (where $\epsilon_1=0.0015$, $\epsilon_2=0.016$), which is approximately 1.529 bits when $K=2$. The gap increases with $K$.\\
$^{**}$The gap per-user is upper bounded by $\frac{1}{2}\log\frac{9(1+\epsilon_1)^2e}{2\pi}+\epsilon_2$ nats, which is approximately 1.006 bits.}
\end{table*}

The remaining part of this paper is organized as follows.
In Sec. II we introduce the OIMAC with power constraints and some useful notations.
Our results for the average-power constrained and for the peak-power constrained OIMAC are provided in Sec. III and Sec. IV, respectively.
Finally, some concluding remarks are given in Sec. V.

\emph{Notation}:
Throughout this paper,
$C$ stands for capacity,
$H(\cdot)$ and $h(\cdot)$ stand for entropy and differential entropy, respectively,
$I(\cdot;\cdot)$ stands for mutual information,
and $I(\mathsf{SNR})$ and $I(\mathsf{PNR})$ stand for the mutual information between $X$ and $X+Z$, $Z\sim{\mathcal N}(0,\sigma^2)$, with respect to SNR and PNR, respectively. We use $\overline{a}$ and $\underline{a}$ to denote upper and lower bounds on a quantity $a$, respectively.
For $i\in\{1,2\}$, we use $\tilde{i}$ to denote the other element in $\{1,2\}$.
The convex closure (or convex hull) of a set of points $\mathcal P$ is denoted by $\mathrm {Conv}(\mathcal P)$.
The asymptotic expression
\begin{equation}
\lim\limits_{t\to\infty}\left[A(t)-B(t)\right]=0
\end{equation}
is denoted as $A(t)\doteq B(t)$.

\section{Optical Intensity Multiple Access Channel}
A discrete-time single-user optical intensity channel with Gaussian noise is given by \cite{LMW09,FH10}
\begin{equation}
\label{DTOIC}
Y=X+Z,
\end{equation}
where $X\ge0$ and $Z\sim{\mathcal N}(0,\sigma^2)$.
A discrete-time two-user OIMAC has two transmitters and one receiver, and the received signal is the linear superposition of the inputs and the noise\cite{KB97,CMA16}:
\begin{equation}
\label{OIMAC}
Y=X_1+X_2+Z,
\end{equation}
where $X_i\ge0, \mspace{4mu}i=1,2$, and $Z\sim {\mathcal N}(0, \sigma^2)$.
This paper considers two types of input power constraints, namely, the per-user average-power constraint as
\begin{equation}
\label{AP}
\mathbf E[X_1]\leq {\mathcal E}_1,\mspace{4mu} \mathbf E[X_2]\leq {\mathcal E}_2,
\end{equation}
and the per-user peak-power constraint as
\begin{equation}
\label{PP}
0\leq X_1\leq {\mathcal A}_1, \mspace{4mu} 0\leq X_2\leq {\mathcal A}_2.
\end{equation}
We define the optical SNR and PNR as $\mathsf{SNR}\triangleq\frac{{\mathcal E}}{\sigma}$ and $\mathsf{PNR}\triangleq\frac{{\mathcal A}}{\sigma}$, respectively,
and denote the SNR and PNR of user $i$ as $\mathsf{SNR}_i$ and $\mathsf{PNR}_i$, respectively.
Throughout the paper, in high-power analysis, we let all $\mathsf{SNR}_i$ or $\mathsf{PNR}_i$ increase simultaneously, i.e., we keep the ratio $\frac{\mathsf{SNR}_i}{\mathsf{SNR}_{\tilde i}}$ or $\frac{\mathsf{PNR}_i}{\mathsf{PNR}_{\tilde i}}$ fixed as input power increases.
These notations and assumptions can be extended directly to a $K$-user OIMAC
\begin{equation}
\label{KOIMAC}
Y=\sum\limits_{i=1}^K X_i+Z.
\end{equation}

In a $K$-user channel, we assume that the users $1,...,K$ send their messages using some codebooks at coding rates $R_1,...,R_K$, respectively, simultaneously.
If all users can decode their messages with vanishing error probabilities as their channel coding lengths grow without bound, we say that the rate tuple $(R_1,...,R_K)$ is achievable (for a formal definition see \cite{NIT}).
For the $K$-user channel, we have the following definitions:
i) The capacity region $\mathcal C$ is the closure of the set of achievable rate tuples $( R_1,...,R_K)$;
ii) The sum capacity is defined as
\begin{equation}
C_\textrm {sum}=\max\left\{\sum\limits_{k=1}^K R_k: (R_1,...,R_K)\in \mathcal C\right\};
\end{equation}
iii) The single-user capacity (or individual capacity) $C_k$ for user $k$ is the supremum of the achievable individual rate for user $k$.

The following single-letter characterization of the capacity region of the OIMAC readily follows from the capacity region of discrete memoryless MAC and
the discretization procedure [\ref{NIT}, Sec. 3.4] (cf. \cite{CAAA}).

\emph{Lemma 1 (Capacity region of OIMAC):
The capacity region of the OIMAC (\ref{OIMAC}) is the convex closure of $\bigcup_{p_{X_1}(x_1)p_{X_2}(x_2)}{\mathcal R}(X_1,X_2)$, where ${\mathcal R}(X_1,X_2)$ is the set of rate pairs $(R_1,R_2)$ satisfying}
\begin{subequations}
\begin{align}
\label{R1}
R_1&\leq I(X_1;Y|X_2),\\
\label{R2}
R_2&\leq I(X_2;Y|X_1),\\
\label{Rsum}
R_1+R_2&\leq I(X_1,X_2;Y),
\end{align}
\end{subequations}
\emph{for a fixed product distribution $p_{X_1}(x_1)p_{X_2}(x_2)$ satisfying the given input constraint.}

However, evaluating this capacity region is difficult since the inputs have a continuous amplitude.
Even for the single-user optical intensity channel, no analytic expression for the capacity is known.
To characterize the capacity region, we will provide outer and inner bounds.

For an additive MAC as (\ref{OIMAC}), two simple but useful facts are given as follows.
The first is
\begin{equation}
\label{sub}
I(X_i;Y|X_{\tilde{i}})=I(X_i;X_i+Z),
\end{equation}
due to
\begin{align}
I(X_i;Y|X_{\tilde{i}})&=h(Y|X_{\tilde{i}})-h(Y|X_1,X_2)\notag\\
&=h(X_i+Z)-h(Z)\notag\\
&=I(X_i;X_i+Z).
\end{align}
The second is
\begin{equation}
\label{add}
I(X_1,X_2;Y)=I(X_1+X_2; Y),
\end{equation}
due to
\begin{align}
I(X_1,X_2;Y)&=h(Y)-h(Y|X_1,X_2)\notag\\
&=h(Y)-h(Z)\notag\\
&=h(Y)-h(Y|X_1+X_2)\notag\\
&=I(X_1+X_2; Y).
\end{align}
These facts help us utilize single-user capacity results in our study on the capacity region of the OIMAC.

\section{Average-Power Constrained OIMAC}

\subsection{Known Single-User Capacity Results}
For an OIMAC with the per-user average-power constraint (\ref{AP}), we utilize results of the single-user optical intensity channel and the additive exponential noise channel \cite{Verdu,AV96,Martinez} to derive capacity bounds.
In this section, we introduce these results.

\emph{Lemma 2 \cite{HK04,LMW09,FH10}:
The capacity of the Gaussian optical intensity channel (\ref{DTOIC}) with an average-power constraint as $\mathbf E[X]\leq{\mathcal E}$ is upper-bounded by}\footnote{The result (\ref{APUB}) was implicitly given in [\ref{HK04}, Sec. III-D], which established an upper bound (Eqn. (21) therein) on effective spectral efficiency of $N$-dimensional time-disjoint signaling in a continuous-time optical intensity channel with Gaussian noise.
For a special case, namely \emph{one-dimensional rectangular} pulse-amplitude modulated (PAM) signaling under an average-power constraint, the discrete-time equivalent channel is exactly the optical intensity channel (\ref{DTOIC}) with an average-power constraint.
By correspondingly substituting $V(\Upsilon_1)=1$ (see [\ref{HK04}, Table I]) and $N=1$ into Eqn. (20) of \cite{HK04} we obtain $C_\textrm{AP-OIC}\leq \log \left(\left(\sqrt T\frac{P}{\sigma}+2\right)\sqrt{\frac{e}{2\pi}}\right)$.
Since the optical power therein is constrained by $\sqrt T P $ (see [\ref{HK04}, Eqn. (9)]), by replacing $\sqrt T P $ with $\mathcal E$ we immediately obtain the result (\ref{APUB}) in Lemma 2.
See \cite{HK04} for more details.}
\begin{equation}
\label{APUB}
C_\textrm{AP-OIC}\leq\frac{1}{2}\log\left(\frac{e}{2\pi}(\mathsf{SNR}+2)^2\right),
\end{equation}
\emph{where $\mathsf{SNR}\triangleq\frac{\mathcal E}{\sigma}$.
The capacity is lower-bounded by}
\begin{subequations}\label{APLB12}
\begin{align}
\label{APLB}
C_\textrm{AP-OIC}&\ge I^\textrm{E}(\mathsf{SNR})\triangleq I(X^\textrm{E};X^\textrm{E}+Z)\\
\label{APLB1}
&\ge\frac{1}{2}\log\left(1+\frac{e}{2\pi}\mathsf{SNR}^2\right),
\end{align}
\end{subequations}
\emph{where} $X^\textrm{E}$ \emph{is an exponential random variable with mean ${\mathcal E}$. The capacity is also lower-bounded by}
\begin{equation}
\label{APLB2}
C_\textrm{AP-OIC}\ge I^\textrm{G}(\mathsf{SNR})\triangleq \max\limits_{\ell>0} I(X^\textrm{G};X^\textrm{G}+Z),
\end{equation}
\emph{where} $X^\textrm{G}$ \emph{is a geometric random variable with mean ${\mathcal E}$ and a probability density function (PDF) parameterized by $\ell$ as}
\begin{equation}
\label{Geo}
p_X(x,\ell)=\sum \limits_{m=0}^{\infty} \frac{\ell}{\ell+\mathcal{E}}\left(\frac{\mathcal{E}}{\ell+\mathcal{E}}\right)^m \delta(x-m\ell), \mspace{10mu} \ell>0,
\end{equation}
\emph{where $\delta(x)$ is the Dirac delta function. At high SNR,}
\begin{equation}
C_\textrm{AP-OIC}\doteq\frac{1}{2}\log\left(\frac{e}{2\pi}\mathsf{SNR}^2\right).
\end{equation}

\emph{Lemma 3\cite{Verdu,AV96}:
The capacity of an additive exponential noise (AEN) channel}
\begin{equation}
\label{AEN}
Y=X+Z, \mspace{4mu}X\ge0, \mspace{4mu} \mathbf E[X]\leq {\mathcal E}_\textrm{s},
\end{equation}
\emph{where $Z$ is an exponential random variable with mean} ${{\mathcal E}_\textrm{n}}$, \emph{is given by }
\begin{equation}
C_\textrm{AEN}=\log\left(1+\frac{{\mathcal E}_\textrm{s}}{{\mathcal E}_\textrm{n}}\right).
\end{equation}
\emph{The PDF of the capacity-achieving input distribution is }
\begin{equation}
\label{mix}
p_X(x)=\frac{{\mathcal E}_\textrm{n}}{{\mathcal E}_\textrm{s}+{\mathcal E}_\textrm{n}}\delta(x)+\frac{{\mathcal E}_\textrm{s}}{\left({\mathcal E}_\textrm{s}+{\mathcal E}_\textrm{n}\right)^2}\exp\left(-\frac{x}{{\mathcal E}_\textrm{s}+{\mathcal E}_\textrm{n}}\right),\mspace{6mu} x\ge 0,
\end{equation}
\emph{and the corresponding output distribution is an exponential distribution with mean} ${\mathcal E}_\textrm{s}+{\mathcal E}_\textrm{n}$.

According to Lemma 3, there exists a probability distribution such that the convolution of its PDF and an exponential PDF
\begin{equation}
\label{Z}
p_Z(z)=\frac{1}{\mathcal E_\mathrm n}\exp\left(-\frac{z}{\mathcal E_\mathrm n}\right)
\end{equation}
is an exponential PDF with mean $\mathcal E_\mathrm s+\mathcal E_\mathrm n$.
That distribution can be simply obtained as follows \cite{Verdu}.
Since the Laplace transform of (\ref{Z}) is
\begin{equation}
\mathbf E\left[\exp(-\mathsf s Z)\right]=\frac{1}{1+\mathcal E \mathsf s},
\end{equation}
to achieve an exponentially distributed output $Y$ in the AEN channel $Y=X+Z$,
the Laplace transform of the PDF of $X$ must be
\begin{equation}
\frac{1+\mathcal E_\mathrm n \mathsf s}{1+(\mathcal E_\mathrm s+\mathcal E_\mathrm n) \mathsf s}=\frac{\mathcal E_\mathrm n}{\mathcal E_\mathrm s+\mathcal E_\mathrm n}+\frac{\mathcal E_\mathrm s}{\mathcal E_\mathrm s+\mathcal E_\mathrm n}\frac{1}{1+(\mathcal E_\mathrm s+\mathcal E_\mathrm n)\mathsf s},
\end{equation}
and the PDF (\ref{mix}) can obtained accordingly.

\subsection{Bounds on Capacity Region of Average-Power Constrained OIMAC}

Our main results for the average-power constrained OIMAC are given in the following two propositions.

\emph{Proposition 1 (Outer bound):
The capacity region of the OIMAC (\ref{OIMAC}) with per-user average-power constraints (\ref{AP}) is outer-bounded by}
\begin{subequations}\label{COB}
\begin{align}
\label{CiOB}
\overline{C}_i&=\frac{1}{2}\log\left({\frac{e}{2\pi}}\left(\mathsf{SNR}_i+2\right)^2\right), \mspace{4mu}i=1,2,\\
\label{CsumOB}
\overline{C}_\textrm{sum}&=\frac{1}{2}\log\left({\frac{e}{2\pi}}\left(\mathsf{SNR}_1+\mathsf{SNR}_2+2\right)^2\right).
\end{align}
\end{subequations}

\begin{IEEEproof}
From (\ref{R1}), the rate of user $i$ must satisfy
$R_i \leq \max\limits_{p_{X_1}(x_1)p_{X_2}(x_2)}I(X_i;Y|X_{\tilde{i}})$.
Combining this with the fact (\ref{sub})
and the single-user capacity upper bound (\ref{APUB}), we obtain (\ref{CiOB}).
From (\ref{Rsum}), the sum rate must satisfy
$R_1+R_2 \leq \max\limits_{p_{X_1}(x_1)p_{X_2}(x_2)}I(X_1,X_2;Y)$.
Combining this with (\ref{add}), by noting that $X_1+X_2$ must satisfy an average-power constraint $\mathbf{E}[X_1+X_2]\leq{\mathcal E}_1+{\mathcal E}_2$, and applying the single-user upper bound (\ref{APUB}), we obtain (\ref{CsumOB}).
\end{IEEEproof}

\emph{Proposition 2 (Inner bound):
The capacity region of the OIMAC (\ref{OIMAC}) with per-user average-power constraints (\ref{AP}) is inner-bounded by a polytope $\mathcal R^{\mathrm{GE}}$ with the following five rate pairs as corner points:}
\begin{align}
\label{R1R2}
(R_1,R_2)=\big\{&(0,0),(I^\textrm{G}(\mathsf{SNR}_1),0),\notag\\
&(I^\textrm{E}(\mathsf{SNR}_1),I^\textrm{E}(\mathsf{SNR}_1+\mathsf{SNR}_2)-I^\textrm{E}(\mathsf{SNR}_1)),\notag\\
&(I^\textrm{E}(\mathsf{SNR}_1+\mathsf{SNR}_2)-I^\textrm{E}(\mathsf{SNR}_2), I^\textrm{E}(\mathsf{SNR}_2)),\notag\\
&(0, I^\textrm{G}(\mathsf{SNR}_2))\big\}.
\end{align}
\emph{A closed-form inner bound consists of all rate pairs $(R_1,R_2)$ such that}
\begin{subequations}\label{RLB}
\begin{align}
\label{RiLB}
R_i&\leq\underline{C}_i=\frac{1}{2}\log\left(1+\frac{e}{2\pi}\mathsf{SNR}_i^2\right), \mspace{4mu}i=1,2,\\
\label{RsumLB}
\mspace{-10mu}R_1+R_2&\leq\underline{C}_\textrm{sum}=\frac{1}{2}\log \left(1+\frac{e}{2\pi}(\mathsf{SNR}_1+\mathsf{SNR}_2)^2\right).
\end{align}
\end{subequations}
\begin{IEEEproof} The achievability of the second and last rate pairs in (\ref{R1R2}) follows directly from Lemma~2.
To prove the achievability of the third and fourth rate pairs, we employ an input distribution $p_{X_1}(x_1)p_{X_2}(x_2)$.
Let $p_{X_i}(x_i)$ be an exponential distribution with mean ${\mathcal E}_i$, and let $p_{X_{\tilde{i}}}(x_{\tilde{i}})$ be as (\ref{mix}) in which we set ${\mathcal E}_\textrm{s}={\mathcal E}_{\tilde{i}}$ and ${\mathcal E}_\textrm{n}={\mathcal E}_i$.
According to Lemma~3, the sum random variable $X_1+X_2$ is exponentially distributed with mean ${\mathcal E}_1+{\mathcal E}_2$.
By combining (\ref{R1}), (\ref{R2}) with (\ref{sub}) we obtain that a rate $R_i= I^\textrm{E}(\mathsf{SNR}_i)$ is achievable for user $i$,
and by combining (\ref{Rsum}) with (\ref{add}) we obtain that a sum rate $R_1+R_2=I^\textrm{E}(\mathsf{SNR}_1+\mathsf{SNR}_2)$ is achievable.
So user $\tilde{i}$ can achieve $R_{\tilde{i}}=I^\textrm{E}(\mathsf{SNR}_1+\mathsf{SNR}_2)-I^\textrm{E}(\mathsf{SNR}_i)$.
Therefore the third and fourth rate pairs in (\ref{R1R2}) are both achievable.
All other rate pairs in the inner bound can be achieved by time sharing \cite{NIT}.

By replacing $I^\textrm{G}(\mathsf{SNR}_1)$ in (\ref{APLB2}) with $I^\textrm{E}(\mathsf{SNR}_1)$, we obtain another achievable rate region $\mathcal R^{\mathrm{E}}$ which is also a polytope with five corner points.
According to (\ref{APLB12}), the region determined by (\ref{RLB}) is a subset of $\mathcal R^{\mathrm{E}}$, and is thus achievable.
\end{IEEEproof}

In \cite{FH10}, it is shown numerically that (\ref{APLB2}) is tighter than (\ref{APLB}) for a wide range of SNR of interest ($-15$ dB to $15$ dB in Fig.~2 therein). However, we have yet to find a proof for $I^\textrm{G}(\mathsf{SNR})\ge I^\textrm{E}(\mathsf{SNR}), \forall \mspace{2mu}\mathsf{SNR}$. Therefore, although we believe that, compared to $\mathcal R^{\mathrm{GE}}$, the inner bound (\ref{RLB}) is always weaker, in the above proof we have to prove it through $\mathcal R^{\mathrm{E}}$ rather than $\mathcal R^{\mathrm{GE}}$.

Combining Proposition~1 and the inner bound (\ref{RLB}) in Proposition~2, we immediately obtain the following corollary.

\emph{Corollary 1}:
\emph{Let $\dot{{\mathcal R}}$ be the achievable rate region given by}
\begin{subequations}
\begin{align}
\label{AR1}
R_i&\leq\dot{C}_i=\frac{1}{2}\log\left({\frac{e}{2\pi}}\mathsf{SNR}_i^2\right), \mspace{4mu}i=1,2,\\
\label{AR2}
R_1+R_2&\leq\dot{C}_\textrm{sum}=\frac{1}{2}\log\left({\frac{e}{2\pi}}(\mathsf{SNR}_1+\mathsf{SNR}_2)^2\right).
\end{align}
\end{subequations}
\emph{Then $\dot{{\mathcal R}}$ approximates the capacity region of the average-power constrained OIMAC (\ref{OIMAC}) to within a vanishing gap as SNR grows without bound.
}

The rate region $\dot{{\mathcal R}}$ is a pentagon as shown in Fig.~1. It is determined by $\dot{C}_i$ and $\dot{C}_\textrm{sum}$, which are high-SNR asymptotic expressions of the single-user capacity and the sum capacity, respectively.
We call $\dot{{\mathcal R}}$ the asymptotic capacity region of the average-power constrained OIMAC.

\begin{figure}[t]
\centering
\includegraphics[width=3.4in,height=2.55in]{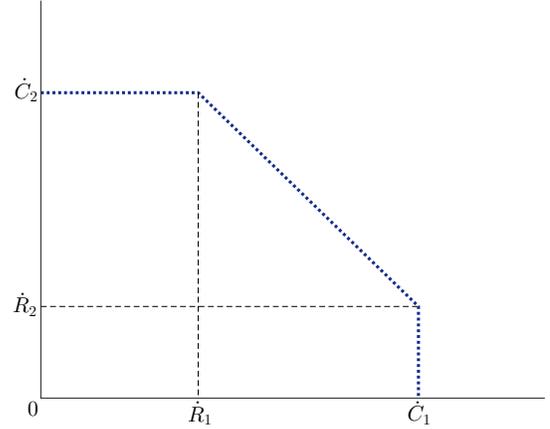}
\caption{An illustration of the asymptotic capacity region of average-power constrained OIMAC.}
\label{0}
\end{figure}
\begin{figure*}[t]
\centering
\includegraphics[width=4.8in,height=3.6in]{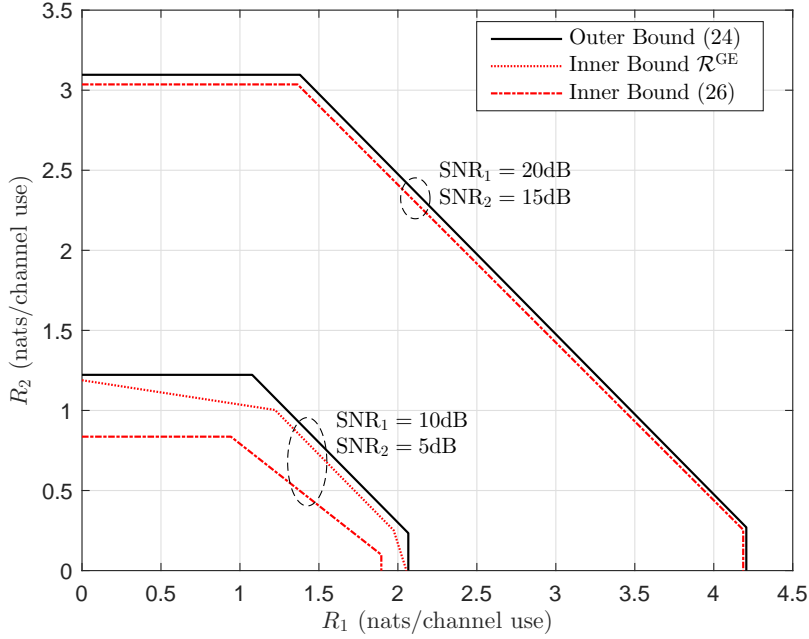}
\caption{Outer and inner bounds on the capacity region of two-user average-power constrained OIMAC.}
\label{1}
\end{figure*}

\emph{Remark 1 (Rate of the second user)}:
Combining (\ref{AR1}) and (\ref{AR2}), we note that when user $i$ asymptotically achieves $\dot{C}_i$, the rate of the second user satisfies
\begin{equation}
\label{Second}
R_{\tilde i}\doteq \dot R_{\tilde i}=\log\left( 1+\frac{\mathsf{SNR}_{\tilde i}}{\mathsf{SNR}_i}\right).
\end{equation}
This can be interpreted as follows.
From Proposition~2, when user $i$ employs an exponential input distribution and achieves the rate $R_i=I^\textrm{E}(\mathsf{SNR}_i)$, which is lower-bounded by the RHS of (\ref{RiLB}),
the other user $\tilde i$, employing an input distribution like (\ref{mix}), can achieve $R_{\tilde i}=I^\textrm{E}(\mathsf{SNR}_1+\mathsf{SNR}_2)-I^\textrm{E}(\mathsf{SNR}_i)$,
which is lower-bounded by\footnote{By (\ref{RLB}) we can show that when user $i$ achieves $\underline{C}_i$ in (\ref{RiLB}), user ${\tilde i}$ can achieve $\underline{C}_\textrm{sum}-\underline{C}_i$, which exceeds $\underline{R}_{\tilde i}$ in (\ref{Rj}). However, the rate pair $(R_i,R_{\tilde{i}})=(I^\textrm{E}(\mathsf{SNR}_i),\underline{C}_\textrm{sum}-\underline{C}_i)$ is not necessarily achievable because $\underline{C}_\textrm{sum}-\underline{C}_i$ may exceed $I^\textrm{E}(\mathsf{SNR}_1+\mathsf{SNR}_2)-I^\textrm{E}(\mathsf{SNR}_i)$.}
\begin{equation}
\label{Rj}
\underline{R}_{\tilde i}=\frac{1}{2}\log \frac{1+\frac{e}{2\pi}({\mathsf{SNR}}_1+{\mathsf{SNR}}_2)^2}{\frac{e}{2\pi}\left({\mathsf{SNR}_i}+2\right)^2}.
\end{equation}
This lower bound can be obtained by
1) combining (\ref{APUB}), (\ref{APLB}), and (\ref{APLB1}) to obtain upper and lower bounds of $I^\textrm{E}(\mathsf{SNR})$,
and 2) applying both bounds as
\begin{equation}
R_{\tilde i}\ge \underline{I}^\textrm{E}(\mathsf{SNR}_1+\mathsf{SNR}_2)-\overline{I}^\textrm{E}(\mathsf{SNR}_i).
\end{equation}
Similarly, an upper bound on $R_{\tilde{i}}$ can be obtained as
\begin{align}
\label{RjU}
{\overline{R}}_{\tilde i}&=\overline{{I}}^\textrm{E}(\mathsf{SNR}_1+\mathsf{SNR}_2)-\underline{{I}}^\textrm{E}(\mathsf{SNR}_i)\notag\\
&=\frac{1}{2}\log \frac{\frac{e}{2\pi}({\mathsf{SNR}}_1+{\mathsf{SNR}}_2+2)^2}{1+\frac{e}{2\pi}{\mathsf{SNR}_i}^2}.
\end{align}
Comparing (\ref{Rj}) and (\ref{RjU}) we obtain (\ref{Second}).
Therefore, the corner points of $\dot{{\mathcal R}}$ are given by
\begin{equation}
(\dot C_i,\dot R_{\tilde i})=\left(\frac{1}{2}\log\left({\frac{e}{2\pi}}\mathsf{SNR}_i^2\right),\log\left( 1+\frac{\mathsf{SNR}_{\tilde i}}{\mathsf{SNR}_i}\right)\right), i=1,2.
\end{equation}

Fig.~\ref{1} shows our capacity bounds for the average-power constrained OIMAC by two examples.
At high SNR, the closed-form inner bound in Proposition~2 is very tight.
At moderate SNR, the closed-form inner bound (\ref{RLB}) becomes looser, but the inner bound $\mathcal R^{\mathrm{GE}}$, which can be evaluated numerically, is still fairly tight.

\begin{figure*}[t]
\begin{align*}
\label{KLBV}
R_{k}=
\begin{cases}
I^\textrm{G}\left(\mathsf{SNR}_{k}\right),&{\mathcal J}=\{k\},\\
I^\textrm{E}\left(\sum\limits_{\tau(\ell)\leq \tau(k)}\mathsf{SNR}_{\ell}\right)-I^\textrm{E}\left(\sum\limits_{\tau(\ell')< \tau(k)}\mathsf{SNR}_{\ell'}\right),&k\in {\mathcal J},\mspace{4mu}{|\mathcal J|}\neq1,\\
0,&k\in {\mathcal J}^{\textrm{c}}.
\end{cases}
\tag{36}
\end{align*}
\end{figure*}
\subsection{Extension to $K$-User OIMAC}
Consider a $K$-user OIMAC as (\ref{KOIMAC}).
Let ${\mathcal K}=\{1,2,...,K\}$, $\mathcal J\subseteq{\mathcal K}$, $|{\mathcal J}|=J\ge 0$ (i.e., ${\mathcal J}=\emptyset$ is allowed), $X_{\mathcal J}=\{X_k:k\in{\mathcal J}\}$, and $R_{\mathcal J}=\sum_{k\in{\mathcal J}}R_k$. Let ${\mathcal J}^{\textrm c}$ denote the complement of ${\mathcal J}$.
By directly extending Lemma~1, we obtain that the capacity region of the $K$-user OIMAC is the convex closure of the rate tuples $(R_1,R_2,...,R_K)$ satisfying
\begin{align}
\label{Rk}
R_{\mathcal J}\leq I(X_{\mathcal J};Y|X_{\mathcal J^{\textrm c}}),\mspace{4mu} \textrm{for all} \mspace{4mu}{\mathcal J}\subseteq{\mathcal K},
\end{align}
for some product distribution $\prod _{k\in{\mathcal K}}p_{X_k}(x_k)$ satisfying the given input power constraints.

Denote the maximum achievable $R_{\mathcal J}$ for all feasible input distributions as $C_{\mathcal J}$.
The following results on the capacity region of the $K$-user OIMAC can be obtained following the same approach in our study on the two-user case.
For brevity we only give outlines of proofs.

\emph{Proposition 3 (Outer bound):
The capacity region of the $K$-user OIMAC (\ref{KOIMAC}) with per-user average-power constraints $\mathbf E[X_k]\leq {\mathcal E}_k, k\in {\mathcal K}$, is outer-bounded by}
\begin{align}
\label{KUB}
\overline{C}_{\mathcal J}= \frac{1}{2}\log\left({\frac{e}{2\pi}}\left(\sum\limits_{k\in{\mathcal J}}\mathsf{SNR}_k+2\right)^2\right),\mspace{4mu}\forall \mathcal J\subseteq{\mathcal K}.
\end{align}

$\mspace{16mu}$\emph{Outline of Proof}:
The proof is similar to that of Proposition 1.
The bound can be derived from (\ref{Rk}) by i) noting that
\begin{align}
\mathbf E\left[\sum\limits_{k\in {\mathcal J}} X_k\right]\leq \sum\limits_{k\in {\mathcal J}} {\mathcal E}_k
\end{align}
must be satisfied $\forall {\mathcal J}\subseteq {\mathcal K}$, and ii) applying the upper bound (\ref{APUB}) to the mutual information $I(X_{\mathcal J};Y|X_{\mathcal J^{\textrm c}})$. \hfill{\small$\blacksquare$}

\emph{Proposition 4 (Inner bound):
Let $\tau$ be a permutation on $\mathcal K$ and $\tau(k)$ be the order of $k$ in $\tau$. For a given ${{\mathcal J}\subseteq {\mathcal K}}$, let ${\mathcal V}_{\mathcal{J}}$ be the set of rate tuples $(R_1,...,R_K)$ satisfying (\ref{KLBV}).}\footnote{When $\ell'$ does not exist (i.e., when $\tau(k)=1$), we let $\mathsf{SNR}_{\ell'}=0$. }
\emph{The capacity region of the $K$-user OIMAC (\ref{KOIMAC}) with per-user average-power constraints $\mathbf E[X_k]\leq {\mathcal E}_k, k\in {\mathcal K}$, is inner-bounded by}
\setcounter{equation}{36}
\begin{equation}
\label{KLB}
{\mathcal R}^K=\mathrm{Conv}\left(\bigcup_{{\mathcal J}\subseteq {\mathcal K}}\mathcal{V}_{\mathcal{J}}\right).
\end{equation}
\emph{A closed-form inner bound slightly weaker than (\ref{KLB}) consists of all rate tuples $(R_1,...,R_K)$ such that}
\begin{align}
\label{KLB1}
R_{\mathcal J}\leq \underline{C}_{\mathcal J}= \frac{1}{2}\log\left(1+{\frac{e}{2\pi}}\left(\sum\limits_{k\in{\mathcal J}}\mathsf{SNR}_k\right)^2\right),\mspace{4mu}\forall \mathcal J\subseteq{\mathcal K}.
\end{align}

\begin{IEEEproof}
An outline of the proof is given in the appendix.
\end{IEEEproof}

The following asymptotic behavior of the capacity region can be obtained by noting that the gap between the upper and lower bounds on $C_{\mathcal J}$ vanishes in the high-SNR limit.

\emph{Corollary 2:
As SNR grows without bound, the capacity region of the $K$-user OIMAC (\ref{KOIMAC}) with per-user average-power constraints $\mathbf E[X_k]\leq {\mathcal E}_k, k\in {\mathcal K}$, can be approximated to within a vanishing gap by the achievable rate region determined by}
\begin{align}
\label{aK}
R_{\mathcal J}\leq \dot{C}_{\mathcal J}=\frac{1}{2}\log\left({\frac{e}{2\pi}}\left(\sum\limits_{k\in{\mathcal J}}\mathsf{SNR}_k\right)^2\right),\mspace{4mu}\forall \mathcal J\subseteq{\mathcal K}.
\end{align}

Note that each of (\ref{KLB1}) and (\ref{aK}) includes $2^K-1$ equations.
The rate regions determined by (\ref{KLB}), (\ref{KLB1}) and the high-SNR rate region determined by (\ref{aK}) are all convex polytopes in the $K$-dimensional space.

\emph{Remark 2}:
According to Proposition~4, a sum rate
\begin{equation}
\max\limits_{(R_1,...,R_K)\in{\mathcal R}^K}\sum\limits_{k\in{\mathcal K}} R_k=I^\textrm{E}\left(\sum_{k\in{\mathcal K}}\mathsf{SNR}_k\right)
\end{equation}
is achievable.
We can also provide an inner bound on the capacity region by
\begin{align}
\label{KLBx}
R_{\mathcal J}\leq I^\textrm{E}\left(\sum\limits_{k\in{\mathcal J}}\mathsf{SNR}_k\right),\mspace{4mu}\forall \mathcal J\subseteq{\mathcal K},
\end{align}
which is tighter than (\ref{KLB1}).
At first glance, the inner bound (\ref{KLBx}) is equivalent to the inner bound (\ref{KLB}) except when $|J|=1$.
However, this is true only at high SNR.
When the SNR is sufficiently low, the rate $I^\textrm{E}(\mathsf{SNR}_1+\mathsf{SNR}_2)-I^\textrm{E}(\mathsf{SNR}_2)$ may exceed $I^\textrm{E}(\mathsf{SNR}_1)$.
In this case the inner bound (\ref{KLBx}) is strictly weaker than (\ref{KLB}), even if we replace $I^\textrm{G}\left(\mathsf{SNR}_{k}\right)$ in (\ref{KLBV}) by $I^\textrm{E}\left(\mathsf{SNR}_{k}\right)$.\footnote{In a Gaussian channel $V=\sqrt{P}U+W$, the mutual information $I(U;V)$ is concave in $P$ (see, e.g., [\ref{GSV05}, Corollary 1]). However, in a Gaussian optical intensity channel $Y=AX+Z$, $X\ge 0$, the mutual information $I(X;Y)$ is not concave in $A$ so that $I^\textrm{E}(\mathsf{SNR}_1+\mathsf{SNR}_2)>I^\textrm{E}(\mathsf{SNR}_1)+I^\textrm{E}(\mathsf{SNR}_2)$ is possible.}

\begin{figure}[t]
\centering
\includegraphics[width=3.4in,height=2.55in]{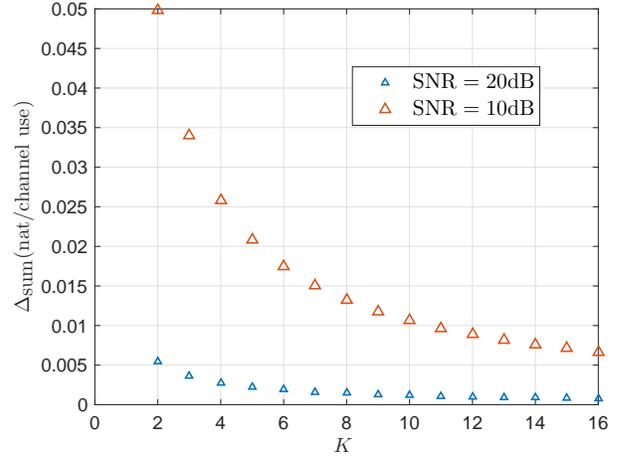}
\caption{Gaps between outer and inner bounds on the sum capacity of OIMAC with symmetric average-power constraint.}
\label{GapK}
\end{figure}

To evaluate the gap between our outer and inner bounds, consider an OIMAC with a symmetric average-power constraint, i.e., ${\mathcal E}_k={\mathcal E}, \forall k\in{\mathcal K}$.
In this case the gap on the sum capacity is
\begin{align}
\Delta_\textrm{sum}(\mathsf{SNR})
&\triangleq \overline{C}_\textrm{sum}-\max\limits_{(R_1,...,R_K)\in{\mathcal R}^K}\sum\limits_{k\in{\mathcal K}} R_k\notag\\
&=\frac{1}{2}\log\left({\frac{e}{2\pi}}\left(K\cdot\mathsf{SNR}+2\right)^2\right)-I^\textrm{E}\left(K\cdot\mathsf{SNR}\right).
\end{align}
This gap is equal to the gap between (\ref{APUB}) and (\ref{APLB}) when their $\mathsf{SNR}$ is scaled by $K$ (cf. [\ref{FH10}, Fig.~2]).
For fixed SNR, as $K$ increases, our bounds on the sum capacity becomes tighter; see Fig.~\ref{GapK}.

\subsection{Discussion on $K$-User OIMAC}

In contrast to the average-power constrained Gaussian MAC, to achieve the capacity region of the average-power constrained OIMAC, different users need to employ different types of input distribution.
Consider the asymptotic capacity region determined by (\ref{aK}).
At high SNR, the boundary of this region includes a face on which the sum rate is maximized (max-sum-rate face).
A corner point of this face is given by
\begin{subequations}
\begin{align}
\label{R1c}
R_1&=\frac{1}{2}\log\left(\frac{e}{2\pi}\mathsf{SNR}_1^2\right),\\
\label{Rkc}
R_{k}&=\log\left(1+\frac{\mathsf{SNR}_k}{\sum_{j<k}\mathsf{SNR}_j}\right),\mspace{4mu}k>1,\mspace{4mu}k\in {\mathcal K},
\end{align}
\end{subequations}
which is asymptotically achieved by employing the input distribution described in the proof of Proposition~4.
That input distribution and the achieved rate for different users, however, is highly asymmetric.
Take, for example, the OIMAC with a symmetric average-power constraint.
In this case, as SNR increases, the rate $R_1$ in (\ref{R1c}) grows without bound, and the corresponding input distribution is an exponential distribution with mean ${\mathcal E}$.
But according to (\ref{Rkc}),
\begin{equation}
R_k= \log\left(1+\frac{1}{k-1}\right), \mspace{4mu}k>1,
\end{equation}
and the corresponding input distribution of the $k$th user has a singleton at zero satisfying $\Pr(X_k=0)=\frac{k-1}{k}$.
If our target is maximizing the sum rate with equal rates for all users (in this case the maximum achievable rate per user is called symmetric capacity \cite{NIT}), then time sharing or rate splitting must be used, while in an average-power constrained Gaussian MAC a single Gaussian input distribution suffices \cite{NIT}.
The symmetric capacity of the OIMAC with a symmetric average-power constraint can also be achieved using time-division multiple access (TDMA) with power (intensity) control \cite{NIT}, which has lower detection complexity than transmitting simultaneously.
However, the optimality of TDMA in terms of sum capacity does not hold if there exists a per-user peak-power constraint (some examples on this fact can be found in \cite{CAAA}).

According to our results, to achieve the sum capacity of the OIMAC with average-power constraint at high SNR, the input distribution for each user must be carefully chosen based on the input power constraints of all users.
A natural question is that if the users still follow a single-user transmission strategy (i.e., employing some near-optimal input distributions for the single-user OIMAC), then how large is the loss on the sum rate?
We give an example to shed some insight on this.
Consider the sum rates achieved by two types of input distributions as follows.
\begin{itemize}
  \item Type I: users follow the asymptotically optimal input distributions given in the proof of Proposition~4.
  \item Type II: the input of each user obeys an exponential distribution with maximum allowed average intensity (asymptotically optimal at high SNR in the single-user case).
\end{itemize}
For simplicity we once again focus on the symmetric average-power constraint ${\mathcal E}_k={\mathcal E}, \forall k\in {\mathcal K}$.
For Type I, the sum of the inputs (we denote it by $S^\mathrm{I}$) is exponentially distributed with mean $K{\mathcal E}$, while for Type II, the sum of the inputs $S^\mathrm{II}=\sum_{k=1}^K X_k$ obeys an Erlang distribution with PDF \cite{GallagerSP}
\begin{equation}
p_S(s)=\frac{s^{K-1}e^{-\frac{s}{\mathcal E}}}{{\mathcal E}^K(K-1)!}, \mspace{4mu}s\ge 0.
\end{equation}
Using the fact
\begin{align}
I(S;\mathsf{SNR}\cdot S+Z)=h&(\mathsf{SNR}\cdot S+Z)-h(\mathsf{SNR}\cdot S+Z|S)\notag\\
=h&(\mathsf{SNR}\cdot S+Z)-h(Z)\notag\\
\doteq h&(\mathsf{SNR}\cdot S),
\end{align}
the high-SNR gap between the sum capacity (asymptotically achieved by Type I) and the sum rate achieved by Type II can be evaluated by the gap between the differential entropies of $S^\mathrm{I}$ and $S^\mathrm{II}$.
The differential entropy of $S^\textrm{I}$ is $\log K e\mathcal E$, and
the differential entropy of $S^\textrm{II}$ is \cite{VR78}
\begin{equation}
h(S^\textrm{II})=\log\left(e^K(K-1)!{\mathcal E}\right)+(1-K)\psi(K),
\end{equation}
where $\psi(\cdot)$ is the digamma function
\begin{equation}
\psi(K)=\sum\limits_{k=1}^{K}{k^{-1}}-\gamma,
\end{equation}
and $\gamma$ is Euler's constant:
\begin{equation}
\gamma\triangleq \lim\limits_{n\to\infty}\left[\sum\limits_{k=1}^{n}k^{-1}-\ln n\right]\approx0.5772.
\end{equation}
Then we obtain
\begin{subequations}
\begin{align}
\label{Delta1}
&\Delta(\mathsf{SNR})\notag\\
&\triangleq[C_{\textrm{sum}}(\mathsf{SNR})-I(S^\textrm{II};S^\textrm{II}+Z)]\\
\label{Delta}
&\doteq(K-1)\psi(K)-\log\left(e^{K-1}(K-1)!K^{-1}\right).
\end{align}
\end{subequations}

\begin{figure}[t]
\begin{minipage}{1\linewidth}
\centering
\includegraphics[width=3.4in,height=2.55in]{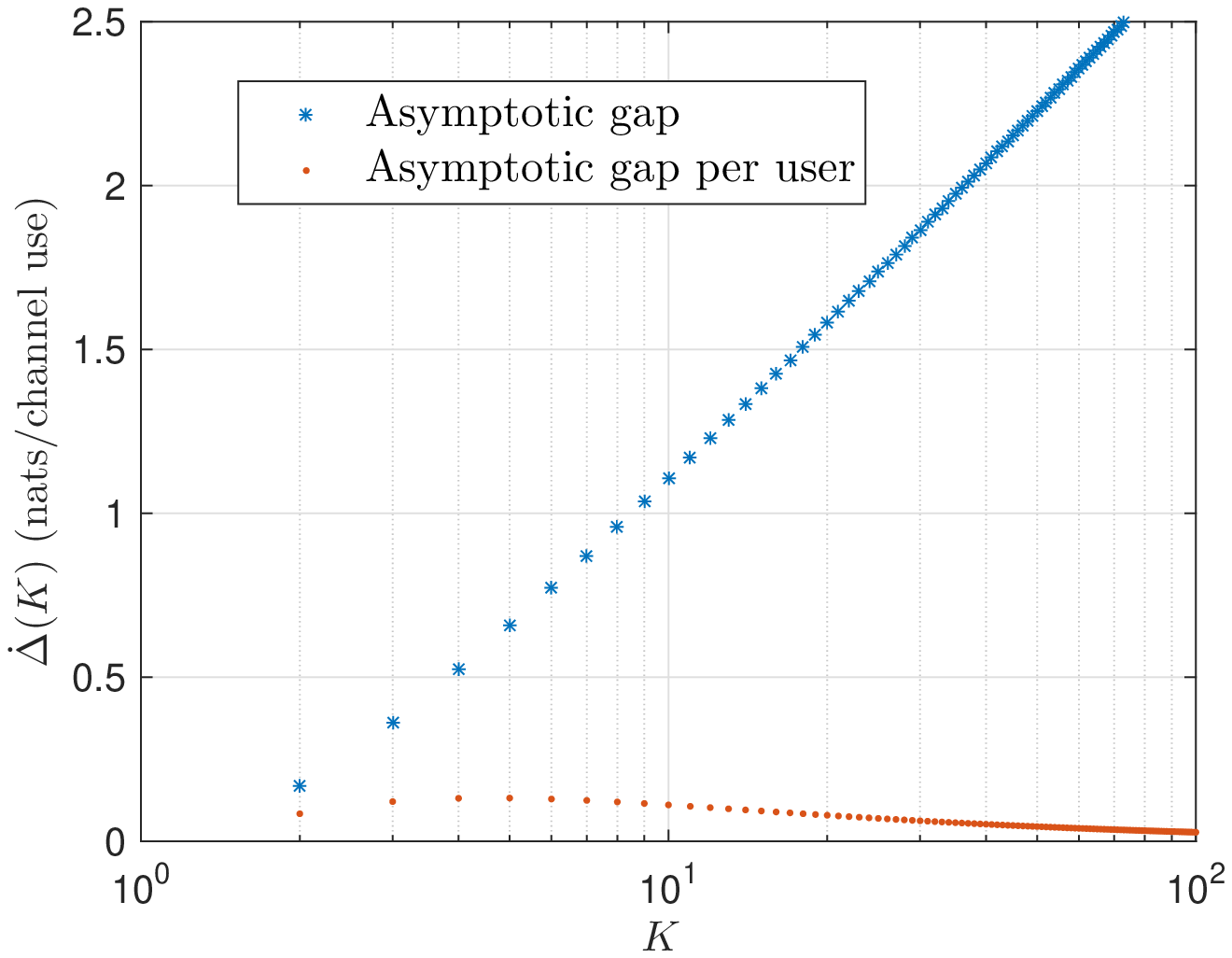}
\caption{Asymptotic gap between sum rates achieved by Type I and Type II input distributions.}
\label{Gap}
\end{minipage}
\begin{minipage}{1\linewidth}
\centering
\includegraphics[width=3.4in,height=2.55in]{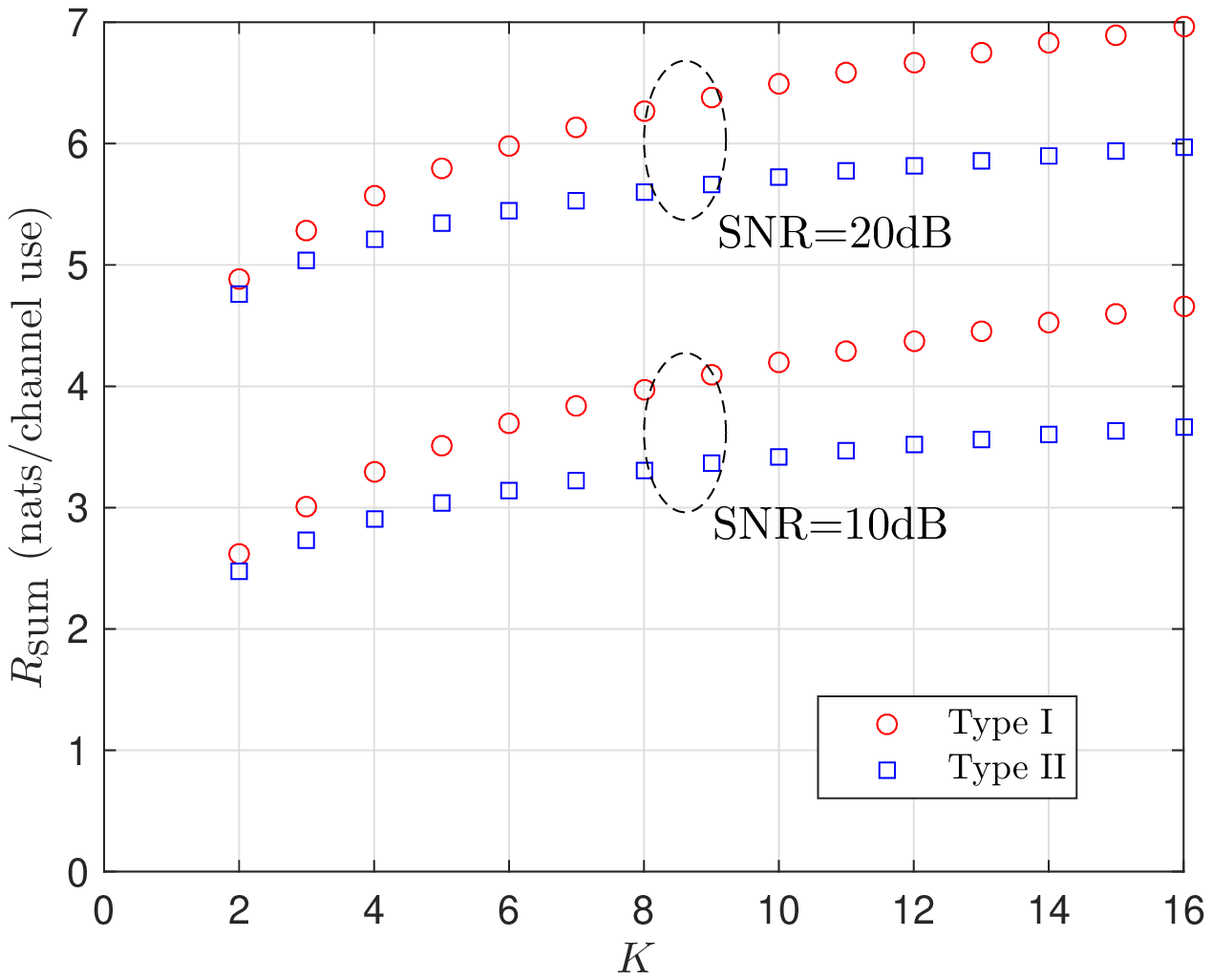}
\caption{Sum rates achieved by Type I and Type II input distributions.}
\label{ExpErl}
\end{minipage}
\end{figure}

In Fig.~\ref{Gap}, the values of (\ref{Delta}), denoted as $\dot{\Delta}(K)$, are plotted.
It is shown that the gap increases linearly as $K$ increases exponentially (in fact $\dot{\Delta}(K)=O(\log K)$).
Therefore, at high SNR, the \emph{per-user} performance loss of Type II input is larger for small $K$, and using a Type-I-like input is more important when there exists a relatively small number of users.
In Fig.~\ref{ExpErl}, by numerically evaluating the input-output mutual information, the sum rates achieved by Type I and Type II inputs are plotted for different numbers of users and finite SNR values.
It is shown that the performance loss of Type II input is more severe when the SNR is lower.

\section{Peak Power Constrained OIMAC}

\subsection{Known Single-User Capacity Results}
For an OIMAC with the per-user peak-power constraint (\ref{PP}), we utilize results of single-user optical intensity channels and certain kinds of peak-power constrained channels to derive capacity bounds.
In this section, we introduce these results.

The following lemma originally describes upper bounds on the capacity of peak-power constrained additive white Gaussian noise (AWGN) channels and was given in \cite{McKellips,TKB}.
Here we have translated the result to optical intensity channels by noting that an optical intensity channel with peak-power constraint ${\mathcal A}$
is equivalent to an AWGN channel with peak-power constraint $|X|\leq\sqrt{P}$ when ${\mathcal A}=2\sqrt{P}$ (i.e., an optical intensity channel with an optical PNR of $\alpha$ dB has the same capacity as a Gaussian channel with an electrical PNR (defined as $P/\sigma^2$) of approximately $\frac{\alpha}{2}+3$ dB).

\emph{Lemma 4:
The capacity of the Gaussian optical intensity channel (\ref{DTOIC}) with a peak-power constraint $0\leq X\leq{\mathcal A}$ is upper-bounded by the McKellips bound \cite{McKellips} as}
\begin{align}
\label{MB}
&C_\textrm{PP-OIC}\leq \overline{C}_\textrm{M}(\mathsf{PNR})\notag\\
&\triangleq\min\left\{\log\left(1+\frac{\mathsf{PNR}}{\sqrt{2\pi e}}\right),\frac{1}{2}\log\left(1+\frac{\mathsf{PNR}^2}{4}\right)\right\}.
\end{align}
\emph{When $\mathsf{PNR}$ satisfies $\frac{1}{2}-{\mathcal Q}(\mathsf{PNR})\ge\frac{\mathsf{PNR}}{\mathsf{PNR}+\sqrt{2\pi e}}$, the capacity is also upper-bounded by \cite{TKB}}
\begin{align}
\label{TKB}
&C_\textrm{PP-OIC}\leq\overline{C}_\textrm{TKB}(\mathsf{PNR})\notag\\
&= H_2\left(\frac{1}{2}-{\mathcal Q}\left(\mathsf{PNR}\right)\right)+\left(\frac{1}{2}-{\mathcal Q}\left(\mathsf{PNR}\right)\right)\log\frac{\mathsf{PNR}}{\sqrt{2\pi e}},
\end{align}
\emph{where $H_2(p)=p\log \frac{1}{p}+(1-p)\log\frac{1}{1-p}$ is the binary entropy function, and ${\mathcal Q}(x)=$ $\frac{1}{\sqrt{2\pi}}\int_x^\infty e^{-\frac{u^2}{2}}\mathrm d u $
is the Q funtion.
The capacity is lower-bounded by \cite{LMW09}}
\begin{align}
\label{PPLB}
C_\textrm{PP-OIC}&\ge I^\textrm{U}(\mathsf{PNR})\triangleq I(X^\textrm{U};X^\textrm{U}+Z)\notag\\
&\ge\frac{1}{2}\log\left(1+\frac{\mathsf{PNR}^2}{2\pi e}\right),
\end{align}
\emph{where} $X^\textrm{U}$ \emph{is a uniformly distributed random variable with support $[0,{\mathcal A}]$.
At high PNR,}
\begin{equation}
C_\textrm{PP-OIC}\doteq\log\frac{\mathsf{PNR}}{\sqrt{2\pi e}}.
\end{equation}

\begin{figure*}[t]
\centering
\includegraphics[width=4.4in,height=3.3in]{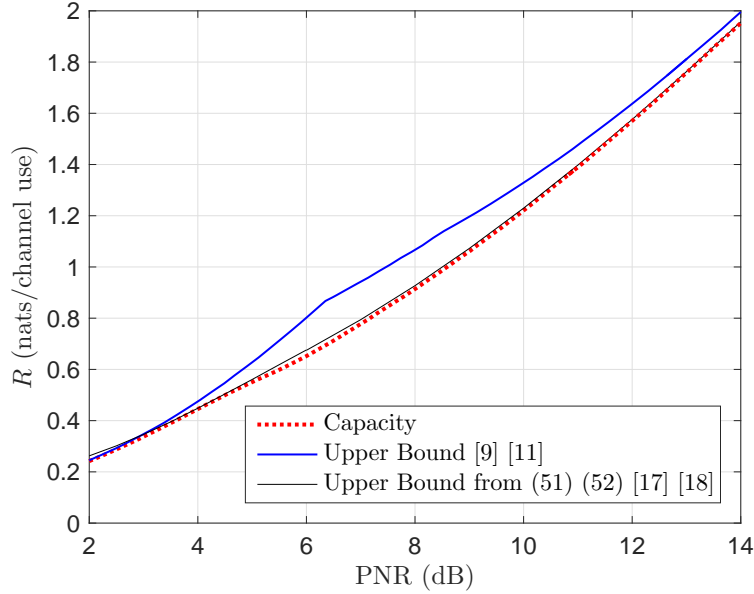}
\caption{Peak-power constrained optical intensity channels: Capacity and its upper bounds.}
\label{M}
\end{figure*}

\emph{Remark 3}: For the peak-power constrained optical intensity channel, several capacity upper bounds have been established in \cite{LMW09,CMA16}, but the upper bounds (\ref{MB}), (\ref{TKB}) in Lemma 4 were not included therein.
In Fig.~\ref{M}, we show an upper bound obtained by combining the upper bounds included in \cite{LMW09,CMA16} (i.e., for each PNR, the shown bound equals to the tightest upper bound therein; see, e.g., [\ref{CMA16}, Fig.~1-(a)]), and another one obtained by combining the upper bounds in Lemma 4.
We also show the channel capacity which can be accurately evaluated with respect to PNR using numerical techniques pioneered in \cite{Smith71}.
It is shown that the bound from Lemma 4 due to \cite{McKellips,TKB} is extremely tight, while the one from \cite{LMW09,CMA16} performs slightly better only when the PNR is below about 3 dB.
Recently, the upper bounds (\ref{MB}) and (\ref{TKB}) have been utilized in the studies on the capacity of optical intensity channels with multiple transmit apertures \cite{MoserMIMO,MoserMIMOA, LiMIMOc}.

\emph{Lemma 5 [\ref{Gallager}, Problem 7.5, pp. 556]:}\footnote{This result has also been noted in German literature in 1960's; see \cite{Oettli} and references therein.}
\emph{Consider an additive noise channel $Y=X+Z$ with input constraint $|X| \leq a$, and $Z$ uniformly distributed over $[-1,1]$.
The capacity of this channel is
}
\begin{equation}
\label{ca}
C(a)=\log(n+1)-(n-a)\log\frac{n+1}{n},
\end{equation}
\emph{where $n=\lceil a\rceil$, and the capacity-achieving input distribution is}
\begin{equation}
\label{DPDF}
p_X(x)=\sum\limits_{m=0}^{n-1}\frac{n-m}{n(n+1)}\left(\delta(a-2m)+\delta(-a+2m)\right).
\end{equation}
\emph{When $a$ is an integer, we have $C(a)=\log(a+1)$ and $p_X(x)=\frac{1}{a+1}\sum\limits_{m=0}^{a}\delta(a-2m)$, which is a discrete uniform distribution.}

\begin{figure}
\begin{minipage}{1\linewidth}
\centering
\includegraphics[width=3.4in,height=2.55in]{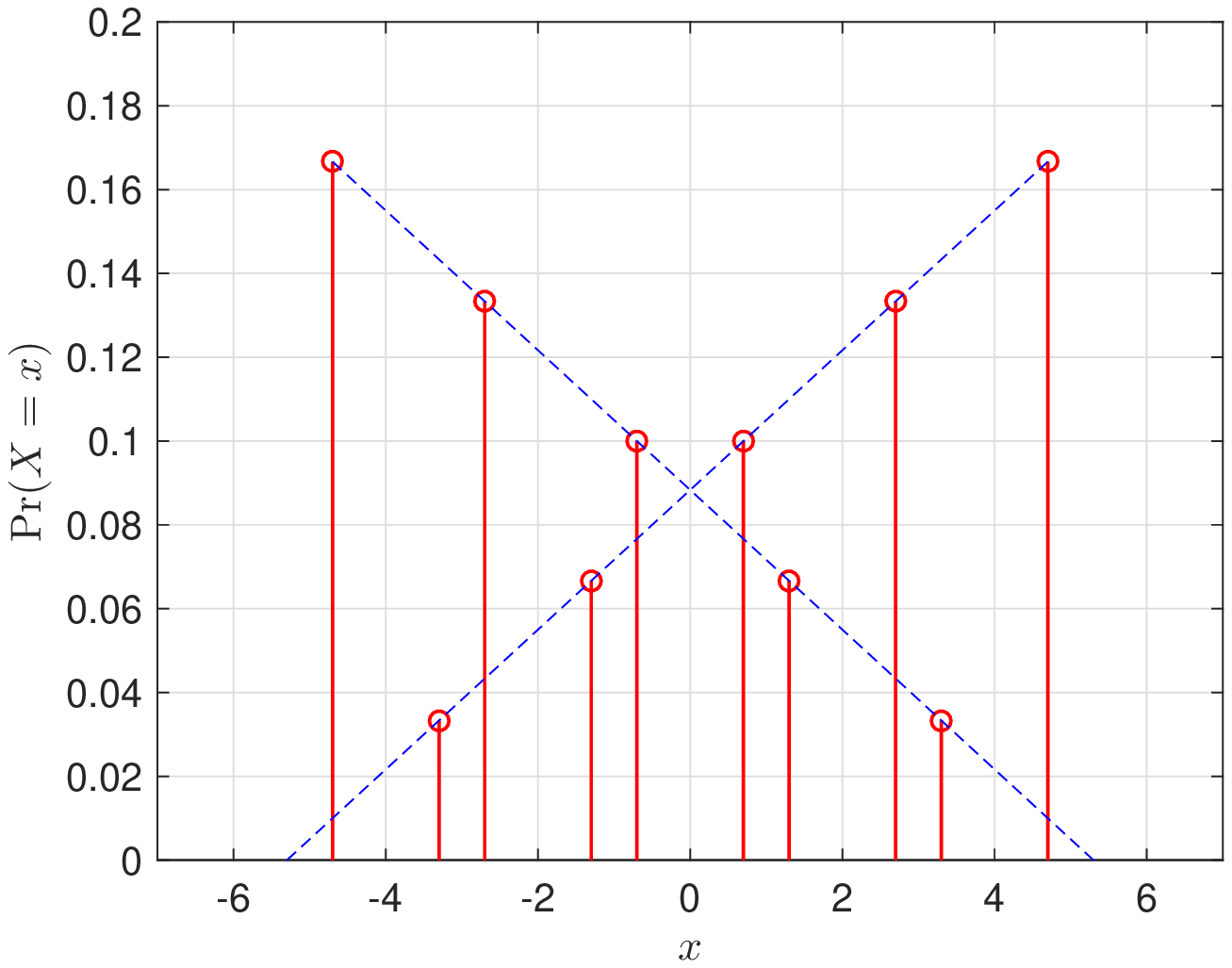}
\caption{Capacity-achieving input distribution in Lemma~5 for $a=4.7$ \cite{Gallager}.}
\label{Gallager1}
\end{minipage}
\begin{minipage}{1\linewidth}
\centering
\includegraphics[width=3.4in,height=2.55in]{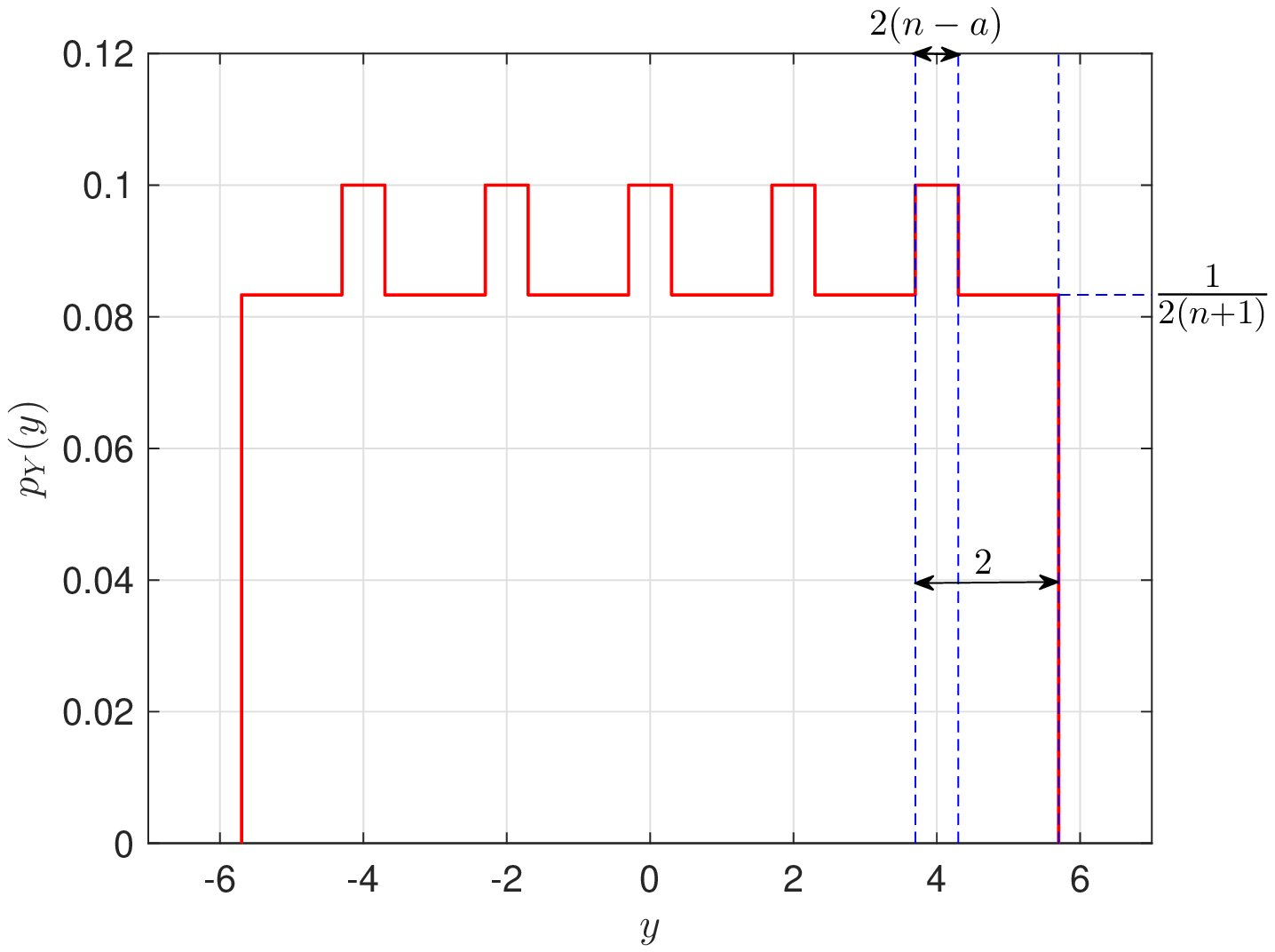}
\caption{Capacity-achieving output distribution in Lemma~5 for $a=4.7$ \cite{Gallager}.}
\label{Gallager2}
\end{minipage}
\end{figure}

We show the capacity-achieving input and output distributions in Lemma~5 for $a=4.7$ in  Fig.~\ref{Gallager1} and Fig.~\ref{Gallager2}, respectively.
\subsection{Bounds on Capacity Region of Peak-Power Constrained OIMAC}

Our main results for the peak-power constrained OIMAC are given in the following two propositions.

\emph{Proposition 5 (Outer bound):
The capacity region of the OIMAC (\ref{OIMAC}) with per-user peak-power constraints (\ref{PP}) is outer-bounded by}
\begin{subequations}\label{PPUB}
\begin{align}
\label{PPUB1}
\overline{C}_i&=\overline{C}(\mathsf{PNR}_i), \mspace{4mu}i=1,2,\\
\label{PPUB2}
\overline{C}_\textrm{sum}
&=\overline{C}(\mathsf{PNR}_1+\mathsf{PNR}_2).
\end{align}
\end{subequations}
\emph{where}
\begin{align}
&\overline{C}(\mathsf{PNR})\notag\\
&\triangleq
\begin{cases}
\min\left\{\overline{C}_\textrm{M}(\mathsf{PNR}),\overline{C}_\textrm{TKB}(\mathsf{PNR})\right\}, &\mathsf{PNR}\leq \mathsf {PNR}^*,\\
\overline{C}_\textrm{M}(\mathsf{PNR}), & \mathsf{PNR}> \mathsf {PNR}^*,
\end{cases}
\end{align}
\emph{where} $\overline{C}_\textrm{M}$ \emph{and} $\overline{C}_\textrm{TKB}$ \emph{are given in (\ref{MB}) and (\ref{TKB}), respectively, and} $\mathsf {PNR}^*$ \emph{is the unique solution of}
\begin{equation}
\frac{1}{2}-{\mathcal Q}(\mathsf{PNR})=\frac{\mathsf{PNR}}{\mathsf{PNR}+\sqrt{2\pi e}},
\end{equation}
\emph{i.e., }$\mathsf {PNR}^*\approx 4.1324$.
\begin{IEEEproof}
This outer bound can be obtained following the same approach as in the proof of Proposition~1.
Specifically, the bound $\overline {C}_i$ is obtained by applying the single-user capacity upper bound in Lemma~4 directly;
The bound $\overline C_\textrm{sum}$ is obtained by $R_\textrm{sum} \leq \max\limits_{p_{X_1}(x_1)p_{X_2}(x_2)}I(X_1+X_2;Y)$, noting that $X_1+X_2\leq {\mathcal A}_1+{\mathcal A}_2$, and applying the single-user capacity upper bound in Lemma~4.
\end{IEEEproof}

We note that the outer bound in Proposition~5 can be further refined by replacing $\overline C(\mathsf{PNR})$ in the right-hand sides of (\ref{PPUB1}) and (\ref{PPUB2}) with $C_\textrm{PP-OIC}(\mathsf{PNR})$, the capacity of the peak-power constrained optical intensity channel, shown in Fig.~\ref{M} (see also Remark 3).
The gain of this refinement is limited, however, because the upper bound in Lemma~4 is already very tight as shown in Fig.~\ref{M}.
Nevertheless, the numerical result of $C_\textrm{PP-OIC}(\mathsf{PNR})$ is indeed helpful for tightening the following inner bound.

\emph{Proposition 6 (Inner bound):
The capacity of the OIMAC (\ref{OIMAC}) with per-user peak-power constraints (\ref{PP}) is inner-bounded by a polytope $\mathcal R^{\mathrm{U}^+}$ with the following five rate pairs as corner points:}
\begin{align}
\label{R1R21}
(R_1,R_2)=\big\{&(0,0), (C_\textrm{PP-OIC}(\mathsf{PNR}_1),0),\notag\\
 &(I^\textrm{U}(\mathsf{PNR}_1),I^{\textrm{U}^+}(a_{2})-I^\textrm{U}(\mathsf{PNR}_1)),\notag\\ &(I^{\textrm{U}^+}(a_{1})-I^\textrm{U}(\mathsf{PNR}_2),I^\textrm{U}(\mathsf{PNR}_2)),\notag\\ &(0, C_\textrm{PP-OIC}(\mathsf{PNR}_2))\big\},
 \end{align}
\emph{where} $C_\textrm{PP-OIC}$ \emph{is the capacity of a single-user optical intensity channel with peak-power constraint; for} $i=1,2$, $a_{i}=\frac{{\mathcal A}_i}{{\mathcal A}_{\tilde i}}$,
\begin{equation}
\label{IU+}
I^{{\textrm{U}^+}}(a_{i})=I(X^\textrm{U}+X^\textrm{D};X^\textrm{U}+X^\textrm{D}+Z),
\end{equation}
\emph{where} $Z\sim{\mathcal N}(0,\sigma^2)$, $X^\textrm{U}$ \emph{is a uniformly distributed random variable with support} $[0, {\mathcal A}_{\tilde i}]$, \emph{and} $X^\textrm{D}$ \emph {has a PDF as}
\begin{equation}
\label{DPDF1}
p_X(x)=\sum\limits_{m=0}^{n_{i}-1}\frac{n_{i}-m}{n_{i}(n_{i}+1)}\left(\delta({\mathcal A}_i-m{\mathcal A}_{\tilde i})+\delta(m{\mathcal A}_{\tilde i})\right),
\end{equation}
\emph{where} $n_{i}=\lceil a_{i}\rceil$.
\emph{A closed-form inner bound weaker than $\mathcal R^{\mathrm{U}^+}$ consists of all rate pairs $(R_1,R_2)$ such that}
\begin{subequations}\label{RLB1}
\begin{align}
\label{RiLB1}
&R_i\leq \underline{C}_i=\frac{1}{2}\log\left(1+\frac{\mathsf{PNR}_i^2}{2\pi e}\right), \mspace{4mu}i=1,2,\\
\label{RsumLB1}
&(\underline{C}_1+\underline{C}_2-\underline{C}_{12})R_1+(\underline{C}_1+\underline{C}_2-\underline{C}_{21})R_2\notag\\
&\leq \underline{C}_1 \underline{C}_{21}+\underline{C}_2 \underline{C}_{12}-\underline{C}_{12}\underline{C}_{21},
\end{align}
\end{subequations}
\emph{where for $i=1,2$,}
\begin{align}
\label{Cii}
\underline{C}_{\tilde i i}=\frac{1}{2}\log\bigg(1+\left(\frac{n_{i}}{n_{i}+1}\right)^{2(n_{i}-a_{i})} \frac{\left(n_{i}+1\right)^2}{2\pi e}\mathsf{PNR}_{\tilde i}^2 \bigg).
\end{align}

\begin{IEEEproof}
The achievability of the second and last rate pairs in (\ref{R1R21}) follows directly from Lemma~4.
To prove the achievability of the third and fourth rate pairs, we employ an input distribution $p_{X_1}(x_1)p_{X_2}(x_2)$, where $p_{X_{\tilde i}}(x_{\tilde i})$
is the PDF of a uniform input distribution over $[0,{\mathcal A}_{\tilde i}]$, and $p_{X_{ i}}(x_{ i})$ is as (\ref{DPDF1}).
By combining (\ref{R1}), (\ref{R2}) with (\ref{sub}),
we obtain that a rate $R_{\tilde i}= I^\textrm{U}(\mathsf{PNR}_{\tilde i})$ for user ${\tilde i}$ is achievable; simultaneously,
by combining (\ref{Rsum}) with (\ref{add}),
we obtain that a sum rate $R_1+R_2=I(X_{\tilde i}^\textrm{U}+X_i^{\textrm D};X_{\tilde i}^\textrm{U}+X_i^{\textrm D}+Z)$ is achievable, which is exactly $I^{\textrm{U}^+}(a_{i})$. So user $i$ can achieve $I^{\textrm{U}^+}(a_{i})-I^\textrm{U}(\mathsf{PNR}_{\tilde i})$, and therefore the third and fourth rate pairs in (\ref{R1R21}) are both achievable.
All other rate pairs in $\mathcal R^{\mathrm{U}^+}$, the convex closure of (\ref{R1R21}), can be achieved using time sharing \cite{NIT}.

To establish the closed-form lower bound (\ref{RLB1}), we first prove that the rate pair $(R_1,R_2)=\left(\underline{C}_1, \underline{C}_{12}- \underline{C}_1\right)$ is achievable.
We employ the input distributions $X^\textrm U$ and $X^\textrm D$ (where we set $i=1$ and $\tilde i=2$) for user 1 and 2, respectively.
Then the achieved sum rate and single-user rate are exactly $I^{{\textrm{U}^+}}(a_{2})$ and $I^\textrm U(\mathsf {PNR}_1)$.
The achievability of $R_1=\underline{C}_1$ can be obtained directly by the single-user capacity lower bound (\ref{PPLB}).
Thus, to show the achievability of $R_2=\underline{C}_{12}- \underline{C}_1$, we only need a proof of
\begin{equation}
\label{IUpLB}
I^{{\textrm{U}^+}}(a_{2})\ge \underline{C}_{12}.
\end{equation}
First, the LHS of (\ref{IUpLB}) can be lower-bounded as
\begin{align}
\label{LHSIUpLB}
&I^{{\textrm{U}^+}}(a_2)\notag\\
&=I(X_1^\textrm{U}+X_2^\textrm{D};X_1^\textrm{U}+X_2^\textrm{D}+Z)\notag\\
&=h(X_1^\textrm{U}+X_2^\textrm{D}+Z)-h(X_1^\textrm{U}+X_2^\textrm{D}+Z|X_1^\textrm{U}+X_2^\textrm{D})\notag\\
&\ge\frac{1}{2}\log\left(\exp \left(2h\left(X_1^\textrm{U}+X_2^\textrm{D}\right)\right)+\exp \left(2h(Z)\right)\right)-h(Z),
\end{align}
where the inequality follows from the entropy power inequality (EPI) \cite{CT06}.
To evaluate $h(X_1^\textrm{U}+X_2^\textrm{D})$, we note that
\begin{equation}
\label{ic}
I(X_2^\textrm{D};X_1^\textrm{U}+X_2^\textrm{D})=C(a_2),
\end{equation}
where the LHS equals to
\begin{align}
\label{hh}
I(X_2^\textrm{D};X_1^\textrm{U}+X_2^\textrm{D})
&=h(X_1^\textrm{U}+X_2^\textrm{D})-h(X_1^\textrm{U}+X_2^\textrm{D}|X_2^\textrm{D})\notag\\
&=h(X_1^\textrm{U}+X_2^\textrm{D})-h(X_1^\textrm{U})\notag\\
&=h(X_1^\textrm{U}+X_2^\textrm{D})-\log {\mathcal A}_1.
\end{align}
Combining (\ref{ic}), (\ref{hh}), and (\ref{ca}) yields
\begin{align}
\label{hud}
&h(X_1^\textrm{U}+X_2^\textrm{D})\notag\\
&=C(a_2)+\log {\mathcal A}_1\notag\\
&=\log\left(n_2+1\right)-\left(n_2-a_2\right)\log\frac{n_2+1}{n_2}+\log{\mathcal A}_1.
\end{align}
Substituting (\ref{hud}) into (\ref{LHSIUpLB}) yields
\begin{align}
I^{{\textrm{U}^+}}(a_{2})
&\ge\frac{1}{2}\log\left(1+\left(\frac{n_2}{n_2+1}\right)^{2(n_2-a_2)}\frac{\left(n_2+1\right)^2}{2\pi e}\mathsf{PNR}_1^2 \right)\notag\\
&=\underline{C}_{12},
\end{align}
and (\ref{IUpLB}) is obtained. So $(R_1,R_2)=\left(\underline{C}_1, \underline{C}_{12}- \underline{C}_1\right)$ is achievable.
By symmetry $\left(\underline{C}_{21}- \underline{C}_2,\underline{C}_2\right)$ is also achievable.
Using time sharing we can further achieve the rate pair
\begin{align}
\label{ts}
&(R_1(\eta),R_2(\eta))\notag\\
&=\big(\eta \underline{C}_1 +(1-\eta)(\underline{C}_{21}-\underline{C}_2),\eta(\underline{C}_{12}-\underline{C}_1)+(1-\eta)\underline{C}_2\big),
\end{align}
where $0\leq\eta\leq 1$.
Note that $(R_1(\eta),R_2(\eta))$ satisfies the linear inequality (\ref{RsumLB1}) with equality, and it also satisfies (\ref{RiLB1}).
Therefore, the set of rate pairs satisfying both $R_1\leq R_1(\eta)$ and $R_2\leq R_2(\eta)$ for all $0\leq \eta\leq 1$, which is an achievable rate region, is equivalent to the set of rate pairs satisfying both (\ref{RiLB1}) and (\ref{RsumLB1}).
This establishes the closed-form inner bound (\ref{RLB1}) in Proposition~6.
\end{IEEEproof}

\subsection{Asymptotic Analysis and Numerical Results}
In this section the tightness of the closed-form inner bound described by (\ref{RLB1}) at high PNR is evaluated.
When $\mathsf{PNR}>\mathsf{PNR}^*$, the outer bound in Proposition~5 can be simplified as
\begin{subequations}
\begin{align}
\overline{C}_i&= \log\left(1+\frac{\mathsf{PNR}_i}{\sqrt{2\pi e}}\right), \mspace{4mu} i=1,2,\\
\overline{C}_\textrm{sum}&= \log\left(1+\frac{\mathsf{PNR}_1+\mathsf{PNR}_2}{\sqrt{2\pi e}}\right).
\end{align}
\end{subequations}
Thus the asymptotic tightness of (\ref{RiLB1}), and also the asymptotic result
\begin{equation}
\label{asySU}
C_i\doteq \log \frac{{\mathsf{PNR}}_i}{\sqrt{2\pi e}},
\end{equation}
can be obtained directly.
Note that $\underline{C}_{{\tilde i}i}$ is a lower bound on the sum capacity (it corresponds to the achievable rate pair $(R_{\tilde i},R_i)=\left(\underline{C}_{\tilde i}, \underline{C}_{\tilde i i}- \underline{C}_{\tilde i}\right)$; see the proof of (\ref{RLB1}).
From
\begin{align}
\overline{C}_{\textrm{sum}}&\doteq\log\frac{(a_i+1){\mathcal A}_{\tilde i}}{\sqrt{2\pi e}\sigma}\notag\\
&=\log \frac{(n_i+1){\mathcal A}_{\tilde i}}{\sqrt{2\pi e}\sigma}+\log\frac{a_i+1}{n_i+1},
\end{align}
and
\begin{align}
\underline{C}_{{\tilde i}i}&\doteq \log\left(\left(\frac{n_i}{n_i+1}\right)^{(n_i-a_i)} \frac{(n_i+1){\mathcal A}_{\tilde i}}{\sqrt{2\pi e} \sigma}\right)\notag\\
&=\log \frac{(n_i+1){\mathcal A}_{\tilde i}}{\sqrt{2\pi e} \sigma}+(n_i-a_i)\log\frac{n_i}{n_i+1},
\end{align}
the asymptotic gap between our upper bound (\ref{PPUB2}) and lower bound (\ref{Cii}) on the sum capacity can be obtained as (without causing ambiguity, indices are omitted hereinafter)
\begin{align}
\Delta(n,\lambda)&=\overline{C}_{\textrm{sum}}-\underline{C}_{{\tilde i}i}\notag\\
&=\log\frac{a+1}{n+1}-(n-a)\log\frac{n}{n+1}\notag\\
&=\log\left(\frac{a+1}{n+1}\left(1+\frac{1}{n}\right)^{n-a}\right)\notag\\
&=\log\left(\left(1-\frac{\lambda}{n+1}\right)\left(1+\frac{1}{n}\right)^\lambda\right),
\end{align}
where $\lambda\triangleq n-a$, which satisfies $0\leq\lambda<1$.
Note that
\begin{align}
\frac{e^{\Delta(n,\lambda)}}{e^{\Delta(n+1,\lambda)}}&=
\frac{\left(1-\frac{\lambda}{n+1}\right)\left(1+\frac{1}{n}\right)^\lambda}{\left(1-\frac{\lambda}{n+2}\right)\left(1+\frac{1}{n+1}\right)^\lambda}\notag\\
&=\frac{n+1-\lambda}{n+2-\lambda}\left(1+\frac{1}{n+1}\right)^{1-\lambda}\left(1+\frac{1}{n}\right)^\lambda\notag\\
&\ge\frac{n+1-\lambda}{n+2-\lambda}\left(1+\frac{1-\lambda}{n+1}\right)\left(1+\frac{\lambda}{n}\right)\notag\\
&=\left(1-\frac{\lambda}{n+1}\right)\left(1+\frac{\lambda}{n}\right)\notag\\
&=1+\frac{\lambda(1-\lambda)}{n(n+1)}\notag\\
&\ge 1,
\end{align}
where the first inequality follows from the fact
\begin{equation}
\left(1+\frac{1}{n}\right)^\lambda\ge 1+\frac{\lambda}{n}.
\end{equation}
Therefore, for a given $\lambda$, the gap $\Delta$ is nonincreasing with $n$.
When $n=1$,
\begin{equation}
\Delta(1,\lambda)=\log\frac{2-\lambda}{2^{1-\lambda}},
\end{equation}
which is continuous, nonnegative, and approaching zero as $\lambda$ tends to zero or one.
Letting $\frac{\mathrm{d}\Delta}{\mathrm{d}\lambda}=0$, we obtain a unique solution $\lambda^*=2-\log_2 e\approx0.5573$.
Therefore, when $\lambda=\lambda^*$ the maximum asymptotic gap of our bounds on sum capacity is achieved, which is
\begin{equation}
\label{max}
\Delta(1,2-\log_2 e)=\log_2\frac{2}{e\ln 2}\approx0.0861 \mspace{4mu}\textrm{bits}.
\end{equation}
This maximum value is achieved when $a_{i}=\frac{{\mathcal A}_i}{{\mathcal A}_{\tilde i}}=\log_2 e-1\approx0.4427$, at the rate pair $(R_{\tilde i},R_i)=(\underline{C}_{\tilde i},\underline{C}_{{\tilde i}i}-\underline{C}_{\tilde i})$.
Since the asymptotic single-user capacity has been found in (\ref{asySU}), by noting that the sum rate of all rate pairs determined by (\ref{ts}) (also (\ref{RsumLB1})) is lower-bounded by $\min\{\underline{C}_{12},\underline{C}_{21}\}$, we can infer that the maximum asymptotic gap (\ref{max}) is also the maximum asymptotic gap between our outer bound (\ref{PPUB}) and inner bound (\ref{RLB1}) on the capacity region.
\begin{figure*}[!t]
\centering
\includegraphics[width=4.8in,height=3.6in]{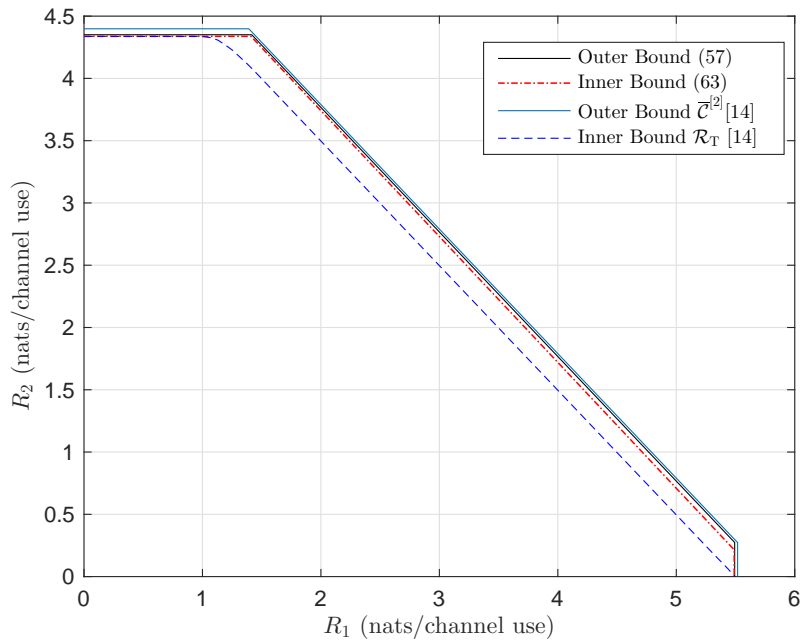}
\caption{Outer and inner bounds on the capacity region of two-user peak-power constrained OIMAC with $\mathsf{PNR}_1=30\textrm{dB}$ and $\mathsf{PNR}_2=25\textrm{dB}$.}
\label{PPh}
\end{figure*}
\begin{figure*}[!t]
\centering
\includegraphics[width=4.8in,height=3.6in]{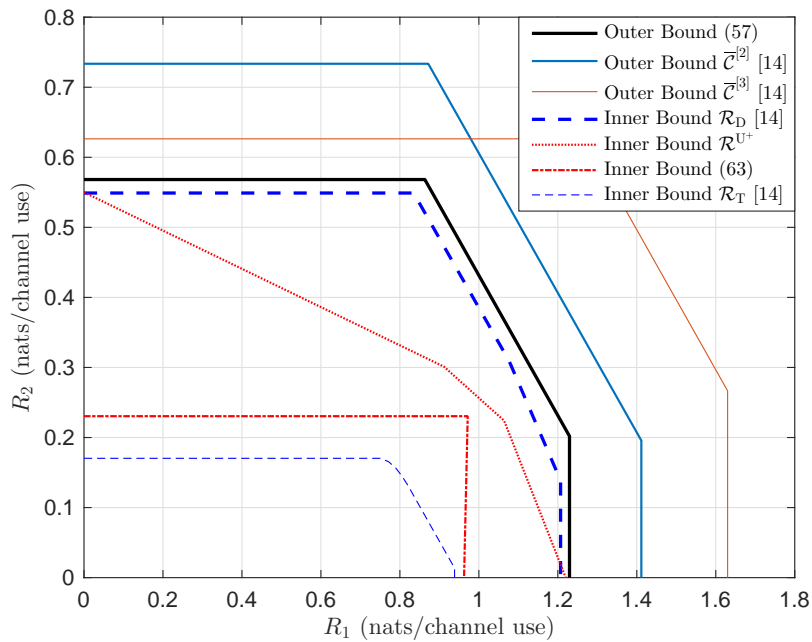}
\caption{Outer and inner bounds on the capacity region of PP-OIMAC with $\mathsf{PNR}_1=10\textrm{dB}$ and $\mathsf{PNR}_2=5\textrm{dB}$.}
\label{PPm}
\end{figure*}
\begin{figure*}[t]
\centering
\includegraphics[width=4.2in,height=3.15in]{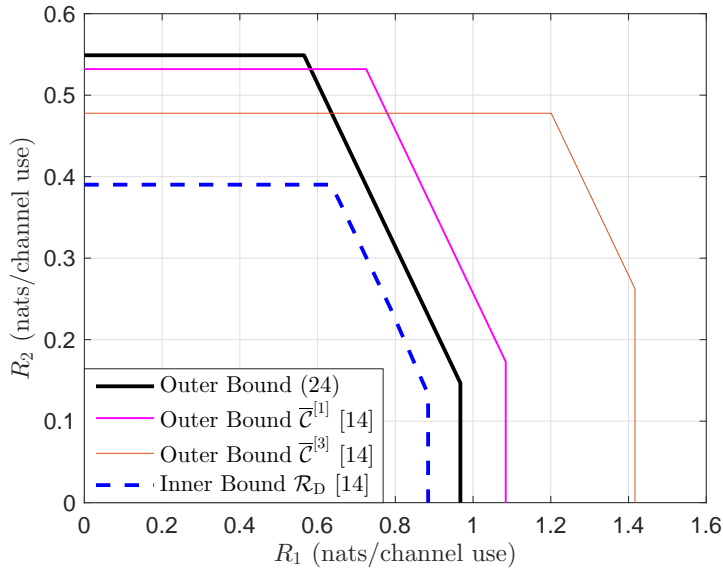}
\caption{The outer bound (\ref{COB}) as an outer bound on the capacity region of two-user average- and peak-power constrained OIMAC with $\mathsf{PNR}_1=10\textrm{dB}$, $\mathsf{PNR}_2=5\textrm{dB}$, and $\frac{\mathcal E_1}{\mathcal A_1}=\frac{\mathcal E_2}{\mathcal A_2}=\frac{1}{5}$.}
\label{APPP}
\end{figure*}

Specifically, for the symmetric case where ${\mathcal A}_1={\mathcal A}_2={\mathcal A}$, we have $\lambda=0$, and the asymptotic gap is $\Delta(1,0)=0$.
So the asymptotic capacity region is determined\footnote{When the input constraint satisfies $\frac{{\mathcal{A}}_i}{{\mathcal{A}}_{\tilde i}}=n_i$, we have $\Delta(n_i,\lambda)=0$, which corresponds to the rate pair $(R_{\tilde i},R_i)=\left(\underline{C}_{\tilde i}, \underline{C}_{\tilde i i}- \underline{C}_{\tilde i}\right)$.
However, the gap $\Delta(n_{\tilde i},\lambda)$ corresponds to the rate pair $(R_i,R_{\tilde i})=\left(\underline{C}_i, \underline{C}_{i\tilde i}- \underline{C}_i\right)$ is not zero since $\frac{{\mathcal{A}}_{\tilde i}}{{\mathcal{A}}_i}=\frac{1}{n_i}$ is not an integer except when $n=1$. In the symmetric case, both gaps vanishes at high PNR and the asymptotic capacity can be determined.} as
\begin{subequations}
\begin{align}
C_i&\doteq \log\frac{\mathsf{PNR}}{\sqrt{2\pi e}},\mspace{4mu}i=1,2\\
C_\textrm{sum}&\doteq \log \frac{2\mathsf{PNR}}{\sqrt{2\pi e}}.
\end{align}
\end{subequations}
Note that in this case exactly
\begin{equation}
C_\textrm{sum}-C_i\doteq 1 \mspace{4mu}\textrm{bit}.
\end{equation}

In summary, in the worst case, we bound the asymptotic capacity region of the peak-power constrained OIMAC to within 0.0861 bits; in the symmetric case, we obtain the asymptotic capacity region.

We next focus on finite-PNR performance. As examples, we show our capacity bounds for the peak-power constrained OIMAC in Fig.~\ref{PPh} and Fig.~\ref{PPm}.
We also compare our results with the bounds in \cite{CAAA} under the ratio $\alpha_i=\frac{{\mathcal E}_i}{{\mathcal A}_i}=\frac{1}{2}$, $i\in\{1,2\}$.
In [\ref{CAAA}, Appendix A], it was shown that in the presence of a peak-power constraint ${\mathcal A}_i$, an average optical power $\mathcal E_i=\mathcal A_i/2$ is sufficient for each user, and using larger average power does not enlarge the capacity region.
This fact implies that the average-power constraint $\mathbf E[X_i]\leq\mathcal E_i=\alpha_i \mathcal A_i$ is redundant for an average- and peak-power constrained OIMAC if $\alpha_i=1/2$, thereby enabling our comparison.

In Fig.~\ref{PPh}, at high PNR, it is shown that our outer and inner bounds almost coincide, and both bounds perform better than the bounds given in [\ref{CAAA}, Fig.~3-a].
The inner bound of \cite{CAAA}, derived by employing a truncated Gaussian input distribution for each user, is fairly close to the outer bounds, but it cannot approach the sum capacity.
In Fig.~\ref{PPm}, at moderate PNR, our inner bound (\ref{RLB1}) is not as tight as in the high-PNR regime.
The inner bound $\mathcal R^{\mathrm{U}^+}$ is tighter, but the inner bound given in [\ref{CAAA}, Fig.~5-a], which is obtained numerically by optimizing a uniformly-spaced discrete distribution, performs best.
Our outer bound is close to this inner bound, thereby bounding the capacity region to within a small gap.
The outer bounds of \cite{CAAA}, however, cannot achieve satisfactory tightness in this example.
Due to the similar bounding techniques, the performances of our outer bounds and those in \cite{CAAA} are determined by the tightness of the single-user upper bounds from which they are derived.
According to the results shown in Fig.~\ref{M}, we can infer that for a wide range of PNRs of interest our outer bound is tighter.

\emph{Remark 4:}
Like in Lemma~4,
all our results in this section can be directly translated to a Gaussian MAC with per-user peak-power constraint as new results.
The peak-power constrained Gaussian MAC has been studied in, e.g., \cite{Verdu86,Ulukus,MMK14}.
In \cite{Verdu86}, it has been shown that if the PNRs are sufficiently small, then the capacity region is achieved by employing equiprobable antipodal signaling with maximum allowable amplitude for each user.
Using this result, numerical examples of the low-power capacity region of the peak-power constrained Gaussian MAC have been provided in \cite{Ulukus}.
In \cite{Ulukus} and \cite{MMK14}, it has been proved that the boundary of the capacity region of the peak-power constrained Gaussian MAC is achieved by discrete input distributions with a finite number of mass points; however, no explicit outer or inner bounds on the capacity region were given therein.

\section{Concluding Remarks}
In this paper, new outer and inner bounds on the capacity region of the OIMAC are established under two types of input power constraints, namely, a per-user average-power constraint, or a per-user peak-power constraint.
We determine the asymptotic capacity region of the average-power constrained OIMAC.
For the peak-power constrained OIMAC, we determine the asymptotic capacity region for the symmetric case, and we bound the asymptotic capacity region to within 0.09 bits in general.
For both cases, at moderate power, several nonasymptotic bounds are found that are also fairly tight.

As we consider either an average- or a peak-power constraint in this paper, a natural future research topic is the case where they are simultaneously constrained.
However, due to the lack of a result like Lemma 3 and Lemma 5,
applying the lower-bounding techniques in this paper to the average- and peak-power constrained case is a nontrivial task.
We leave this for future studies.
Additionally, we remark that for some range of parameters the outer bound derived in this paper may refine the ones in \cite{CAAA} derived for the average- and peak-power constrained OIMAC; see Fig. \ref{APPP} for an example.

Finally, it is worthwhile to note that in recent years there has been a lot of research activity in communications over multiple-input and multiple-output (MIMO) optical intensity channels where the user may employ multiple transmit apertures and/or multiple detectors \cite{EMH,BSTVH,Zeng,FH13}.
Several works have investigated the channel capacity of MIMO optical intensity channels \cite{CRAMIMOh, CRAMIMOl, MoserMIMO,MoserMIMOA, LiMIMOc}.
Similar to the OIMAC, in a MIMO optical intensity channel a detector observes a superposition of signals from multiple transmit apertures.
However, in the OIMAC the devices of different users cannot cooperate, while in a MIMO optical intensity channel a user can send a vector input in one channel use by its multiple transmit apertures, and jointly observe the received signal by its multiple detectors (if any).
Such a fundamental difference results in different capacity bounding techniques.

\begin{appendices}
      \section*{APPENDIX\\Outline of Proof of Proposition~4}

The proof follows the same pattern as the proof of Proposition~2.
First we establish the achivability of ${\mathcal V}_{\mathcal K}$, which is the set of all $K!$ corner points of the max-sum-rate face of ${\mathcal R}^K$.
Based on the first step, we then establish the achivability of $\bigcup\limits_{{\mathcal J}\subseteq {\mathcal K}}\mathcal{V}_{\mathcal{J}}$, which includes all corner points of ${\mathcal R}^K$.
Finally, all rate tuples in ${\mathcal R}^K$ can be achieved using time sharing.
Some details are given as follows.

Consider a permutation $\tau$ on $K$.
We omit the indices of users, and alternatively index a user by its order: $X_k$ is denoted as $X_{(m)}$ if $\tau(k)=m$.
We employ an input distribution $\prod\limits_{m\in {\mathcal K}}p_{X_{(m)}}\left(x_{(m)}\right)$, and let
i) $p_{X_{(1)}}\left(x_{(1)}\right)$ be an exponential distribution with mean ${\mathcal E}_{(1)}$, and
ii) for every $1<m\leq K$, $p_{X_{(m)}}\left(x_{(m)}\right)$ be as (\ref{mix}) in which
\begin{align}
{\mathcal E}_\textrm{s}={\mathcal E}_{(m)},\mspace{8mu} {\mathcal E}_\textrm{n}={\sum\limits_{\ell\in{{\mathcal L}_{(m)}}}{{\mathcal E}_{(\ell)}}},
\end{align}
where ${\mathcal L}_{(m)}=\{(\ell):\ell\in{\mathcal K},1\leq \ell <m\}$.
In this setting, for every ${\mathcal L}_{(m)}\subset\mathcal K $, $\sum_{\ell\in {\mathcal L}_{(m)}} {{X_{(\ell)}}}$ and $X_{(m)}+\sum_{\ell\in {\mathcal L}_{(m)}} {{X_{(\ell)}}}$ are both exponentially distributed with mean $\sum_{\ell\in{\mathcal S}_{(m)}} {{{\mathcal E}_{(\ell)}}}$ and ${\mathcal E} _m+\sum_{\ell\in{\mathcal S}_{(m)}} {{{\mathcal E}_{(\ell)}}}$, respectively.
Following the same approach as that in the proof of Proposition~2, we obtain an achievable rate tuple $\left(R_{(1)},...,R_{(K)}\right)$ which is determined by
\begin{equation}
\label{RLm}
R_{{\mathcal L}_{(m)}}=I^\textrm{E}\left(\sum\limits_{m\in{\mathcal L}_{(m+1)}}\mathsf{SNR}_m\right),\mspace{4mu} m\in{\mathcal K}.
\end{equation}
Note that when ${\mathcal J}={\mathcal K}$, (\ref{RLm}) is equivalent to (\ref{KLBV}).
Therefore, by considering all $K!$ different permutations on $\mathcal K$ we can establish the achievability of ${\mathcal V}_{\mathcal K}$, which further
guarantees the achievability of ${\mathcal V}_{\mathcal J}$ except when $|{\mathcal J}|=1$.
The achievability of ${\mathcal V}_{\mathcal J}, \mspace{4mu}{\mathcal J}=\{k\}$ follows directly from Lemma~2.
The rate region
${\mathcal R}_K$ is achieved using time sharing between rate tuples in $\bigcup\limits_{{\mathcal J}\subseteq {\mathcal K}}\mathcal{V}_{\mathcal{J}}$. 

According to (\ref{APLB12}), the rate region determined by (\ref{KLB1}) is a subset of ${\mathcal R}_K$ and it is thus achievable.
\end{appendices}

\section*{Acknowledgements}

The authors would like to thank the anonymous reviewers for their careful review and helpful comments.

\begin{IEEEbiography}
[{\includegraphics[width=1in,height=1.25in,clip,keepaspectratio]{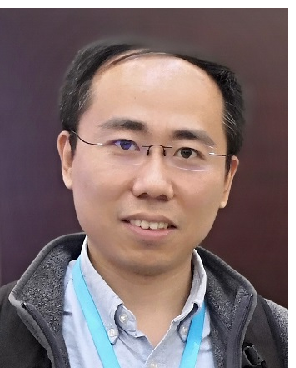}}]
{Jing Zhou} received the Ph.D. degree from the Beijing University of Posts and Telecommunications, Beijing, China, in 2013.
He is currently a Post-Doctoral Research Associate with the Department of Electronic Engineering and Information Science, University of Science and Technology of
China.
His research interest includes digital communications, information theory and applications, and optical wireless communications.
\end{IEEEbiography}

\begin{IEEEbiography}
[{\includegraphics[width=1in,height=1.25in,clip,keepaspectratio]{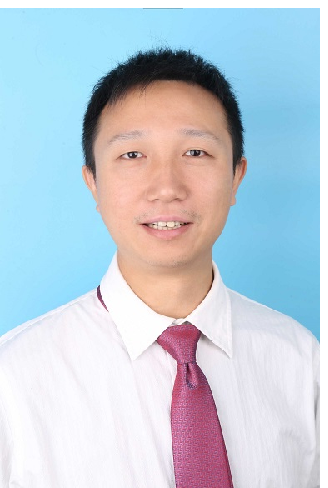}}]
{Wenyi Zhang} (S'00-M'07-SM'11) is currently a professor with the Department of Electronic Engineering and Information Science, University of Science and Technology of China. He received his bachelor¡¯s degree in automation from Tsinghua University in 2001, and his master¡¯s and Ph.D. degrees in electrical engineering from University of Notre Dame, in 2003 and 2006, respectively. He was affiliated with the Communication Science Institute, University of Southern California, as a Post-Doctoral Research Associate, and with Qualcomm Incorporated, Corporate Research and Development. His research interest includes wireless communications and networking, information theory, and statistical signal processing. He was an editor for IEEE Communications Letters, and is currently an editor for IEEE Transactions on Wireless Communications.
\end{IEEEbiography}

\end{document}